*Statement of Significance

**Statement of significance**

The present research work is focused in finding a preventive treatment of bone infection based on Mesoporous Bioactive Glasses (MBGs) with antibacterial adhesion properties obtained by zwitterionic surface modification. MBGs exhibit unique nanostructural, textural and bioactive characteristics. The novelty and originality of this manuscript is based on the design and optimization of a straightforward functionalization method capable of providing MBGs with zwitterionic surfaces that are able to inhibit bacterial adhesion without affecting their cytocompatibility. This new characteristic enhanced the MBG properties to avoid the bacterial adherence onto the implant surfaces for bone tissue engineering applications. Subsequently, it could help to decrease the infection rates after implantation surgery, which represents one of the most serious complications associated to surgical treatments of bone diseases and fractures.

Manuscript submitted to

# Prevention of Bacterial Adhesion to Zwitterionic Biocompatible Mesoporous Glasses

S. Sánchez-Salcedo, A. García, María Vallet-Regí*

Departamento de Química Inorgánica y Bioinorgánica. Facultad de Farmacia. Universidad Complutense de Madrid. Plaza Ramón y Cajal s/n. 28040 Madrid. Spain. Networking Research Center on Bioengineering, Biomaterials and Nanomedicine (CIBER-BBN), Madrid, Spain.

\* Corresponding author: Fax: +34 394 1786; Tel.: +34 91 394 1843; E-mail address: vallet@ucm.es (M. Vallet-Regí)

The two first authors contributed equally to this work.

**Abstract**

Novel materials, based on Mesoporous Bioactive Glasses (MBGs) in the ternary system $SiO_2$-$CaO$-$P_2O_5$, decorated with (3-aminopropyl)triethoxysilane (APTES) and subsequently with amino acid Lysine (Lys), by post-grafting method on the external surface of the glasses (named MBG-$NH_2$ and MBG-Lys), are reported. The surface functionalization with organic groups did not damage the mesoporous network and their structural and textural properties were also preserved despite the high solubility of MBG matrices. The incorporation of Lys confers a zwitterionic nature to these MBG materials due to the presence of adjacent amine and carboxylic groups in the external surface. At physiologic pH, this coexistence of basic amine and carboxilic acid groups from anchored Lys provided zero surface charge named zwitterionic effect. This behaviour could give rise to potential applications of antibacterial adhesion. Therefore, in order to assess the influence of zwitterionic nature in *in vitro* bacterial adhesion, studies were carried out with *Staphylococcus aureus*. It was demonstrated that the efficient interaction of these zwitterionic pairs onto the MBG surfaces reduced bacterial adhesion up to 99.9% compared to bare MBGs. In order to test the suitability of zwitterionic MBGs materials as bone grafts, their cytocompatibility was investigated *in vitro* with MC3T3-E1 preosteoblasts. These findings suggested that the proposed surface functionalization strategy provided MBG materials with notable antibacterial adhesion properties, hence making these materials promising candidates for local bone infection therapy.

## 1. Introduction

The so called third generation biomaterials can be directly classified in the field of bone tissue engineering. Since 2004, when Chen *et al*. prepared for the first time highly ordered Mesoporous Bioactive Glasses (MBGs) with superior *in vitro* bone-forming bioactivities[1], MBGs based in the ternary system $SiO_2$-$CaO$-$P_2O_5$ have been widely investigated and regarded as optimum candidates for lost bone regeneration. Moreover, the ordered mesoporous arrangement, with high surfaces and pore volume, allows using MBGs in drug delivery systems for the treatment of bone tissue diseases [2, 3]. In addition, the high amount of free silanol groups present on MBG surfaces allows to open an interesting route to the anchoring of new species through a covalent attachment, thus improving the applications of these materials[4,5].

Post-operative implant infections are one of the most serious complications associated with surgical treatments of bone diseases and fractures with bone grafts and prostheses[6,7]. In the issue of osteomyelitis (an inflammatory process that leads to bone destruction), the bacteria of biggest concern include *Staphylococcus aureus* and *Staphylococcus epidermidis*[8,9]. Bacteria typically secrete polymeric materials after their association to form protective coatings known as biofilms. Biofilms have been defined as "aggregates of microorganisms in which cells are frequently embedded in a self-produced matrix of extracellular polymeric substances that are adherent to each other and/or a surface" [10]. The biofilm further impedes the activity of the host defenses and/or antibiotic therapy, requiring surgical intervention to remove the implant as the only effective option. According to the National Healthcare Safety Network (NHSN) between 2006 and 2008, the postoperative infection rates associated with orthopaedic surgery range from 1.11 to 3.36% for open reduction of fracture, from 0.58 to 1.60% for knee replacement and from 0.67 to 2.40% for hip replacement, depending on patient risk. Applying these rates to the total amount of hip and knee replacements performed in the United States results in an estimated 6,000-20,000 events of surgical site infection per year, associated with hip and knee replacements alone[11,12]. More recently, NHSN reported a 44.2% of surgical site infections by *S. aureus* in open reduction of fracture, hip prosthesis and knee prosthesis among others between 2011 and 2014[13]. This problem brings serious financial consequences for both, patients and healthcare provider, because the most effective treatment is the implant removal by surgical intervention, with the subsequent cost increase[14].

Zwitterionic polymers, such as poly(carboxybetaine methacrylate) and poly(sulfobetaine methacrylate), containing quaternary ammonium as a positive charge and carboxylate and sulfate as negative charges, have been reported to be good ultralow fouling materials[15]. These zwitterionic-based materials are receiving great attention due to their effectiveness, robustness, and stability[16,17]. Recently, zwitterionization of biomaterials has emerged as a groundbreaking strategy to endow surfaces with high resistance against non-specific protein adsorption, bacterial adhesion and/or biofilm formation for dentistry or orthopaedics applications[18,19]. Vallet-Regí research group have reported a straightforward way to synthesize mesoporous silica ceramics exhibiting zwitterionic surfaces that are able to inhibit bacterial adhesion, whereas allowing osteoblast adhesion and proliferation under *in vitro* conditions[7,20-22]. In this work, a novel scientific effort is devoted to provide bioactive bioceramics such as MBGs with antibacterial adhesion capability while preserving their biocompatible behaviour.

Zwitterionic species are characterized by having an equal number of positively and negatively charged groups within a molecule, hence maintaining overall electrical neutrality. It can bind water molecules even more strongly than hydrophilic materials via electrostatically induced hydration, becoming an important part in achieving interfacial bioadhesion resistance[23,24]. Here we present MBGs with zwitterionic functionalization through amino acid incorporation to the surface. Amino acids are well-known natural zwitterions compounds, having a carboxyl group (-COOH) and an amine group ($-NH_2$) with a third reactive group linkable with MBG surfaces. Several authors have reported the usefulness of amino acid as low fouling strategies to avoid bacterial adhesion[15,25].

Herein, we report for the first time the design and synthesis of a new nanostructured zwitterionic MBGs with antibiofouling capability that inhibits bacterial adhesion. The cytocompatibility of non-zwitterionic and zwitterionic MBG materials was investigated *in vitro* with MC3T3-E1 preosteoblasts cells.

## 2. Materials and Methods

All chemicals were purchased from Sigma-Aldrich, except (3-aminopropyl)triethoxysilane (APTES, 97% wt) that was purchased by ABCR, and used as received without further purification. All surface functionalization reactions were carried out under an inert atmosphere using Schlenk techniques and anhydride solvents.

### 2.1 75/85-MBGs synthesis.

Ordered mesoporous materials in the ternary $SiO_2$-CaO-$P_2O_5$ system were obtained as described in ref.[26]. Briefly, the MBGs were synthesized by using evaporation induced self-assembly (EISA)[27] method and non-ionic surfactant $EO_{20}PO_{70}EO_{20}$ (Pluronic P123) as structure directing agent. Tetraethyl orthosilicate (TEOS), triethyl phosphate (TEP), and calcium nitrate tetrahydrate ($Ca(NO_3)_2 \cdot 4H_2O$) were used as $SiO_2$, $P_2O_5$ and CaO sources, respectively. Two different compositions were synthesized, denoted as 85-MBG and 75-MBG. The amount of reactants and nominal compositions are collected in Table 1. The resulting transparent membranes were calcined at 700°C in air for 6 hours to obtain the finally 75-MBG and 85-MBG powders. Finally, the powders were sieved below the 63 µm. The final compositions of 75-MBG and 85-MBG samples were calculated by X-ray fluorescence (XRF) (Table 1).

## 2.2 75/85-MBG surface functionalization with (3-aminopropyl)triethoxysilane (APTES)

75-MBG and 85-MBG samples were surface functionalized directly post-synthesis with amine groups using (3-aminopropyl)triethoxysilane (APTES). A sample of 2 g of 75/85-MBG was dried and degassed overnight at 110ºC. Afterward, 75/85-MBG was dispersed in 20 mL anhydrous toluene and the suspension was stabilized at $N_2$ atmosphere. Subsequently, a colorless solution formed by 7.2 mL (30.8 mmol) APTES and 25 mL dry toluene was added dropwise, with constant stirring, to a suspension of 75/85-MBG. The mixture was refluxed 24 h at 80ºC under $N_2$ atmosphere. Finally the solid was filtered and washed with copious amount of toluene and water. The white solid was dried at 120°C for 48 h. The resulting amine surface functionalized sample was denoted as 75-MBG-$NH_2$ and 85-MBG-$NH_2$, respectively.

## 2.3 Lysine functionalization of aminated 75/85-MBG-$NH_2$ samples

To provide the 75/85-MBG surfaces with zwitterionic nature, the materials were post-synthesis functionalized with Lysine (Lys) through glutaraldehyde (GA) linkage to amine groups. 1 g of 75/85-MBG-$NH_2$ was suspended in a solution of 2.5 mL of GA 50% w/v and 47.5 mL of deoxygenated water in a 100 mL round-bottom flask under an inert atmosphere to prevent oxidation. The mixture was stirred 1 hour at room temperature and an orange solid coloration was observed. The solid product was filtered and washed several times with water and subsequently added to a Lys-NHBoc aqueous solution (0.13 mg, (0.53 mmol), in 62.5 mL of deoxygenated water). The suspension was stirred overnight at room temperature. The orange solid was collected by filtration and washed exhaustively with water and then dried under vacuum. Finally, to check out Lys terminal amine, the orange solid was treated with trifluoroacetic acid (0.5 mL TFA in 20 mL dry dichloromethane) for 3 hours. The $CH_2Cl_2$ was removed by evaporation and the orange solid was suspended in water to wash by centrifuging several times with water, and finally dried at 30°C under vacuum for two days. The final orange solids were denoted as 75-MBG-Lys and 85-MBG-Lys, respectively. 30 mg of dried of all MBG powders were conformed in a 6 mm diameter and 1 mm height disks using 2.75 MPa uniaxial pressure.

## 2.4. Characterization of MBGs materials

Chemical compositions of MBGs were determined by X-ray fluorescence (XRF) spectroscopy, using a Philips PANalytical AXIOS spectrometer (Philips Electronics NV), with X-rays generated by the RhK$\alpha$ line at $\lambda = 0.614$ Å. The XRF-analyzed compositions are listed in Table 1. The percentage of surface functionalization with APTES and later Lys incorporation through GA linkage was carried out by thermogravimetric analyses (TG). TG measurements were carried out under a dynamic air atmosphere between 25 and 950ºC with a heating rate of 5ºC/min using a Perkin-ElmerDiamond analyzer (Perkin-Elmer,USA). Surface functionalization was studied by solid state magic angle spinning nuclear magnetic resonance (MAS NMR) and cross polarization magic angle spinning nuclear magnetic resonance (CP MAS NMR). The $^{13}C$, $^{29}Si$ and $^{31}P$ spectra were obtained on a Bruker Avance AV-400WB spectrometer equipped with a solid state probe using a 4 mm zirconia rotor and spun at 12 kHz for $^{13}C$, 10 kHz for $^{29}Si$ and 6 kHz in the case of $^{31}P$. The structural characteristics of the resulting materials were determined by powdered X-ray diffraction (XRD) in a Philips X'Pert diffractometer equipped with CuK$\alpha$ (40 kV, 20 mA). Electron microscopy was carried out using a JEOL 2100 electron microscope operating at 200 kV and equipped with an Oxford Link EDX probe. TEM images were recorded using a CCD camera (MultiScan model 794, Gatan). The textural properties of samples were obtained by nitrogen adsorption/desorption analyses at 77K on a Micromeritics ASAP 2020 instrument (Micromeritics Co, Norcross, USA). The surface area ($S_{BET}$) was determined using the Brunauer-Emmett-Teller (BET) method[28]. The total pore volume ($V_T$) was calculated from the amount of $N_2$ adsorbed at a relative pressure of 0.97. The average mesopore diameter was obtained from the adsorption branch of the isotherm by means of the Barret-Joyner-Halenda (BJH) method[29]. FTIR spectra were collected in a Thermo Nicolet Nexus equipped with a Goldengate attenuated total reflectance (ATR) device. ζ-potential measurements were carried out to determine the pH conditions under which the zwitterionic nature of the material surfaces were preserved in aqueous medium, *i.e.* the isoelectric point (IEP), which was closely related to the zero point charge[30]. The pH was adjusted by adding appropriate volumes of 0.10 M HCl or 0.10 M KOH solutions and ζ-potential measurements were performed in a Zetasizer Nano Series instrument from Malvern Instruments Ltd. (UK).

**2.5. Cell culture tests**

The MBG disks were sterilized under ultraviolet light for 7 min each side[31]. After that, disks were stabilized in complete medium that consist on: α-Modified Eagle's Medium (α-MEM, Sigma Chemical Company, St. Louis, MO, USA) with 10% fetal bovine serum (FBS, Gibco, BRL), 1 mM L-glutamine, penicillin/streptomycin (400 mg/mL BioWhittaker Europe, Belgium), under a $CO_2$ (5%) atmosphere at 37°C for 24 h in 24-well tissue culture plates (Cultek SLU, Spain). Complete medium was supplemented with β-glycerolphosphate (50 mg, Life Technologies SA, Spain) and L-ascorbic acid (Life Technologies SA, Spain) for ALP test. Cell culture experiments were performed using the well-characterized mouse preosteoblastic cell line MC3T3-E1 (subclone 4, CRL-2593; ATCC, Mannassas, VA). Controls in tissue culture plastic in the absence of samples were always carried out.

**2.5.1 Cellular adhesion and spreading assays.**

Cell morphology was studied into disks using an Eclipse TS100 inverted optical microscope (Nikon) after 24 hours. Fluorescence microscopy was also carried out for the observation of attached cells onto the disks. After fixed and permeabilized samples, they were incubated with Atto 565-conjugated phalloidin (dilution 1:40, Molecular Probes) which stains actin filaments. Then, samples were washed with PBS and the cell nuclei were stained with 1 M diamino-20-phenylindole in PBS (DAPI) (Molecular Probes). Fluorescence microscopy was performed with an EVOS FL Cell Imaging System equipped with tree Led Lights Cubes ($\lambda_{EX}$ (nm); $\lambda_{EM}$ (nm)): DAPI (357/44; 447/60), RFP (531/40; 593/40) from AMG (Advance Microscopy Group).

**2.5.2 Cell viability**

MC3T3-E1 cells were plated onto the different samples at a density of $10^4$ cells/mL of osteogenic medium consisting of complete medium for 2 and 4 days at 37ºC in a humidified atmosphere of 5% $CO_2$. Cell growth was analysed using MTS (3-(4,5-dimethylthiazol-2-yl)-5-(3-carboxymethoxyphenyl)-2-(4-sulfophenyl)-2H-tetrazolium) assay. The MTS reduction assay was performed using a commercial assay and following the manufacturer's protocol (CellTiter Aqueous One Solution Cell Proliferation Assay). The absorbance at 490 nm was then measured in a Unicam UV-500 UV-visible spectrophotometer (Thermo Spectronic, Cambridge, UK).

**2.5.3 Cytotoxicity assay: Lactate dehydrogenase (LDH) activity.**

LDH activity released from the preoteoblasts cells was considered for cell injury measurement. This indicates the degree of cytotoxicity caused by the test sample. The assay is based on the reduction of nicotinamide adenine dinucleotide (NAD) by LDH.

The resulting colored compound is measured spectrophotometrically at 340 nm. The measurements were made at 24 and 48 h of seeding by using a commercial kit (Spinreac, Spain)[31].

### 2.5.4 Cell differentiation

Cell differentiation was evaluated by alkaline phosphatase (ALP) activity. The osteoblast-like phenotype was assessed by measuring ALP activity of MC3T3-E1 cells grown on zwitterionic and non-zwitterionic disks. ALP activity was measured after 4d using the Reddi and Huggins method[32] with commercially kit (Spinreact, Spain).

### 2.6 Bacterial adhesion assays

These preliminary studies were conducted to determine bacterial adhesion to non-zwitterionic and zwitterionic MBG disks by quantification of adhered bacteria test. Gram-positive *S. aureus* (ATCC 29213) was used as model bacteria. Prior to the antibacterial growth assay, samples were sterilized ultraviolet light for 7 min each side followed by soaking in 1 mL Todd-Hewitt Broth (THB) medium (Sigma- Aldrich, USA) at 37ºC for 2 h in order to stabilize samples for the antibacterial assay. The macropore size distributions of the resulting pieces were measured by Hg intrusion porosimetry in the $5 \cdot 10^{-3}$-$3 \cdot 10^{2}$ μm range using a AutoPore IV 9500 porosimeter (Micromeritics Co., Norcross, USA). *S. aureus* bacteria were grown to mid-logarithmic phase in THB at 37ºC under orbital stirring at 100 rpm until the optimal density as measured at 600 nm reached 1.0. At this point the bacteria from culture were collected by centrifugation (Labofuge 400 centrifuge, Thermo Scientific, USA) at 1500 rpm for 10 min at room temperature, washed three times with sterile PBS, pH 7.4 (Sigma-Aldrich, USA) and subsequently re-suspended into a concentration of $1.3 \times 10^{5}$ cells/mL under orbital stirring at 100 rpm for 90 min[33-35].

### 2.6.1 Quantification of adhered bacteria.

With the aim of quantifying the adhered bacteria, each piece was placed in a 1.5 mL Eppendorf vial (Nirco, Spain) containing 1 mL of sterile PBS, followed by 30 s sonication in a low power bath sonicator (Selecta, Spain). This sonication process was performed three times, assuming that 99.9% of adherent bacteria were removed[36]. Then 100 µL of each sonication product was cultivated on Tryptic Soy Agar (TSA) (Sigma Aldrich, USA) plates, followed by incubation at 37ºC overnight. Determination of the number of colony-forming units (CFU) resulting from the overall sum of the three sonication processes allowed determination of the number of bacteria originally adherent on the pieces. The resulting values were normalized as a function of the geometric surface of the disk pieces. The number of CFUs per $cm^2$ was calculated by normalization of the counted CFUs as a function of the surface area of the disk-shaped pieces in the 1-600 µm range. This normalization process has been carried out by taking into account the results derived from Hg intrusion porosimetry of the materials. Thus, *S. aureus*, which exhibit sizes of around 1 µm, are unable to penetrate the mesoporous structure (3-6 nm) but they are able to penetrate into int ergr anular po res ($\geq$ 1 µm). This fact was confirmed by SEM studies of the materials surfaces after bacterial adhesion assay, as described below.

**2.6.2 Morphological studies by scanning electron microscopy**

The surface characterization of samples after 90 min bacterial incubation was performed by SEM in a JEOL model JSM-6400 microscope. Before the SEM studies the attached bacteria were fixed with 2.5 vol.% GA (50 wt.%, Sigma-Aldrich, USA) in PBS, pH 7.4 and dehydrated by slow water replacement using a series of graded ethanol solutions (35%, 50%, 70%, and 90%) in deionized water, with a final dehydration step in absolute ethanol before critical point drying (Balzers CPD 030, Liechtenstein). The materials were mounted on stubs and gold plated in vacuum using a sputter coater (Balzers CPD 030, Liechtenstein)[37].

## 2.7 Statistics

Cytocompatibility and antibacterial assays data obtained are expressed as means standard deviations of the independent experiments indicated in each case. Statistical analysis were performed using the Statistical Package for the Social Sciences (SPSS). Statistical comparisons were made by analysis of variance (ANOVA). The Scheffé test was used for post hoc evaluation of differences between groups. In all statistical evaluations, $p < 0.05$ was considered as statistically significant.

## 3. Results and discussion

### 3.1 Characterization of non-zwitterionic and zwitterionic MBGs

Ordered mesoporous materials in the ternary $SiO_2$-$CaO$-$P_2O_5$ system were easily prepared in alcoholic medium under acidic conditions by using a non-ionic surfactant Pluronic® P123 as structure-directing agent and evaporation induced self-assembly (EISA) method. Tetraethyl orthosilicate (TEOS), triethyl phosphate (TEP), and calcium nitrate tetrahydrate ($Ca(NO_3)_2 \cdot 4H_2O$) were used as $SiO_2$, $P_2O_5$ and $CaO$ sources, respectively (Table 1). The weight percentages of $SiO_2$, $P_2O_5$ and $CaO$ of the resulting 75/85-MBG materials after being calcined were determined by X-ray fluorescence (XRF) measurements. Two different $CaO$ contents were prepared and the nominal and experimental compositions were in agreement (Table 1).

Lysine (Lys) is an amino acid that was selected to functionalize the surface of MBGs for one main reason. Lys provides amine and carboxylic groups (zwitterionic nature) in the same molecule which is a great advantage for synthetic purposes: MBG materials containing the same number of cationic and anionic terminal groups in their external surfaces. Post-synthesis method was selected due to it involves an organic modification process, once the free-surfactant MBGs have been already obtained. The organic grafting on the silica surface leads to a more heterogeneous distribution of the organic functional groups than co-condensation method[38]. It is reasonable to presume that, after surface decorating, the organic functional groups tightly congregate on the surface of the material outer of mesoporo walls[5].

Scheme 1 shows a summary of the surface modification strategy followed in the present work. In a first step, 75/85-MBG samples were functionalized following a post-synthetic grafting method with an excess of functionalizing agent, (3-aminopropyl)triethoxysilane (APTES). A condensation reaction between organic functional alcoxysilane and silanol groups of 75/85-MBG systems was placed generating a covalent anchorage. The second part of the synthesis consisted in the anchorage of amine N-protected Lysine (Lys-NHBoc) through GA linkage to amine groups present in 75/85-MBG-$NH_2$ materials (GA acts as a linker between terminal amino groups[39,40] provided of aminosilane APTES and free -$NH_2$ group of amino acid protected Lys-NHBoc) under aqueous contitions. Finally, removal of the Boc protecting group by treatment with trifluoroacetic acid (TFA) in dichloromethane[41,42] produced the 75/85-MBG-Lys samples with amino and carboxylic adjacent groups that provide zwitterionic nature. Table 2 summarizes some of the most relevant features of the systems synthesized in this work. The amount of aminosilane anchored in the surface 75/85-MBG-$NH_2$ samples were determined by thermogravimetric (TG) analyses. The results revealed that 75-MBG-$NH_2$ and 85-MBG-$NH_2$ samples exhibit comparable behaviour, with weight losses *ca*. 5% up to 600ºC. With the incorporation of GA and Lys the percentage of organic matter content increased. This was in accordance with a greater weight losses obtained by TG measurements as it can be seen in Figure S1 (Supporting Information). The different grade of Lys anchorage observed between 75-MBG-Lys and 85-MBG-Lys samples could be due to amine groups' loss during GA and Lys-NHBoc incorporation processes in aqueous media. This behaviour could be due to the samples with high amount of calcium inside the framework (*e.g.* 75-MBG-NH$_2$) are more soluble than 85-MBG-NH$_2$ in aqueous media[42]. So, 75-MBG-NH$_2$ sample had less amine groups available to Lys incorporation through GA linkage.

Structural characterization by powder X-ray diffraction (XRD) revealed that almost all synthesized samples exhibit ordered mesoporous arrangements (Figure 1A). The starting materials, 75/85-MBG, displayed a sharp diffraction maxima at 2θ in the region of 1.0-1.4º, assigned to the (10) reflection, along with a poorly resolved peaks at around 2.0 and 2.3º that can be assigned to the (11) and (20) reflections. These maxima were indexed on the basis of an ordered two-dimensional (2D) hexagonal structure (plane group *p6mm*)[26,43]. The intensity of the XRD (10) maximum was constant indicating that mesoporous structure has been maintained after successively functionalization of 85-MBG sample to obtain 85-MBG-NH$_2$ and 85-MBG-Lys, respectively. However, XRD patterns of 75-MBG-NH$_2$ and 75-MBG-Lys samples showed a progressive loss of mesopore arrangement through successive surface functionalization steps with less reaction yield. Transmission electron microscopy (TEM) studies of zwitterionic samples 75/85-MBG-Lys confirmed the results obtained by XRD. Figure 1B (left) displayed a TEM image that corresponded to wormlike mesoporous 75-MBG-Lys sample. Figure 1B (right) corresponded to TEM images of 85-MBG-Lys sample obtained with the electron beam parallel and perpendicular to the mesoporous channels axis show "well ordered" hexagonally arranged mesostructures.

Textural properties of the MBG mesoporous materials have been performed by N$_2$ adsorption/desorption analyses (Figure 2 and Table 2). The isotherms can be identified as type IV according to the IUPAC classification, which are typical of mesoporous solids[44]. The presence of H1 type hysteresis loops in the mesopore range indicated the existence of open ended cylindrical mesopores with narrow pore size distributions, which are characteristic of MBGs[29,45]. The surface area ($S_{BET}$) experienced a decrease from 75/85-MBG to 75/85-MBG-NH$_2$ which is in agreement with the covalent anchorage of aminosilane with silanol groups of 75/85-MBG systems. The subsequent Lys incorporation maintained similar values of this parameter. Pore volume and pore diameter displayed a progressive decrease from 75/85-MBG to 75/85-MBG-NH$_2$ and 75/85-MBG-Lys which is in agreement with the appropriate successively functionalization of MBG materials and, as was expected, the wall thickness ($t_{wall}$) increased after organic functionalization.

In order to evaluate the covalent attachment of the different organic frameworks to the surface of 75/85-MBG precursor samples the $^{13}$C and $^{29}$Si single pulse (SP) and cross-polarization (CP) magic-angle spinning (MAS NMR) solid state spectra were collected. Figure 3 shows the $^1$H→$^{13}$C CP/MAS NMR spectra of 75/85-MBG-NH$_2$ and 75/85-MBG-Lys samples, with the proposed structure and corresponding peak assignments. Due to the absence of material organic present in the initial materials there were no signals in the $^1$H→$^{13}$C CP/MAS NMR spectra of the 75-MBG and 85-MBG samples (data not showed). The $^1$H→$^{13}$C CP/MAS NMR spectra of 75-MBG-NH$_2$ and 85-MBG-NH$_2$ samples, showed in Figure 3 (red), provided clear evidence that they were indeed functionalized as intended. Both spectra showed similar pattern with three prominent signals at *ca.* 44 ppm (C3: methylene group bonded to the terminal amine group *C*H$_2$NH$_2$), 25 ppm (C2: central methylene group, CH$_2$*C*H$_2$CH$_2$), and 10 ppm (C1: the methylene group bonded to the silicon atom, *C*H$_2$Si). In addition, 75-MBG-NH$_2$ spectrum displayed two small signals around 63 and 18 ppm assigned to ethoxy groups ($CH_3CH_2$OSi) from incomplete hydrolysis of APTES during the first functionalization step. The $^1H \rightarrow ^{13}C$ CP/MAS NMR spectra of zwitterionic 75/85-MBG-Lys samples exhibited signals of carbons corresponding to GA linker (C4/4´, C5/5´ and C6) and amino acid Lys extreme (carbons numbered from C7 to C12). It is worth to notice the characteristic presence of a signal around 150 ppm assigned to imine groups (C4 and C4´, N=$C$H). These signals indicated the correct anchorage between amino terminal groups from APTES and Lys-NHBoc through a GA linkage[39]. In addition, no signal at *ca.* 200 ppm was observed from aldehydes, indicating complete linkage between $NH_2$-terminal groups. The successful zwitterionic process was confirmed by the appearance of the typical signals of amino acid at approximately 175 ppm (C12, $C$=OOH) and 60 ppm (C11, $C$-$NH_2$), assigned to carbon of carboxylic acid group and carbon near to amine group from Lys, respectively. Unfortunately, the *tert*-butyl signals from the Boc protective group overlap with the signals of the APTES moiety and other aliphatic signals (C6-10, $C$H_2 from the linker GA and Lys moieties), appearing as broad peaks. Thus, it is difficult to determine only by $^{13}C$ NMR the total deprotection of the amino groups after treatment with TFA. With the aim to clarify the deprotection FTIR spectroscopy were carried out to compare Boc-protected and deprotected materials (Figure S2, Supporting Information). The absence of signal assigned to ester group ($\nu$ (C=O)$_{ester}$ ≈ 1745 $cm^{-1}$) in 75/85-MBG-Lys samples could confirm the deprotection of the zwitterionic materials. Moreover, zeta potential ($\zeta$) measurements showed changes in the surface charge depending on the pH (data see below). In general, the surface charge becomes more positive after Boc deprotection, which is in agreement with the presence of free Lys -$NH_2$/-$NH_3^+$ groups. In addition, the results obtained by TG measurements (Figure S1, Supporting Information) revealed that there were slightly different weight losses between materials before and after Boc deprotection steps. Thus, the successful immobilization of Lys on the mesoporous materials has been confirmed.

$^{29}$Si NMR spectroscopy was used to evaluate the network connectivity of MBGs as a function of different percentage of $SiO_2$ and CaO in the chemical composition. Further analysis of the surface functionalization of MBG materials was made by $^{29}$Si MAS NMR spectroscopy. Figure 4 (left) compares the quantitative spectra from the direct polarization method (SP) obtained for the bare MBG materials (75/85-MBG) with those obtained for the aminosilane functionalized (75/85-MBG-NH$_2$) and amino acid incorporated materials (75/85-MBG-Lys). Table 3 summarizes all the chemical shifts, populations of these silicon environments calculated by Gaussian line-shape deconvolutions peak areas, silica network connectivity and condensation degree for each composition. All the $^{29}$Si SP MAS NMR spectra displayed resonances at around -92, -101 and -111 ppm that represent $Q^2$ [(NBO)$_2$-*Si*-(OSi)$_2$], $Q^3$ [(NBO)-*Si*-(OSi)$_3$] and $Q^4$ [*Si*(OSi)$_4$] silicon sites, respectively (NBO = non-bonding oxygen)[46,47]. Preparing ordered mesoporous glasses in a ternary system is not obvious. $SiO_2$-CaO-$P_2O_5$ contains both network formers ($SiO_2$ and $P_2O_5$) and a modifier (CaO) that disrupts the mesophase formation during the self-assembling of surfactant and inorganic species[43,48]. The lower connectivity index and the lower $Q^4$ peak areas observed in 75-MBG compared to 85-MBG sample confirm the role of $Ca^{2+}$ as network modifiers (75-MBG has higher CaO content (*ca.* 20% molar composition) than 85-MBG (*ca.* 10% molar composition)). When the $^1$H→$^{29}$Si CP MAS NMR spectra were studied $Q^2$ signals were divided into two different signals associated with different types of silicon, related to hydrogen ($Q^2_H$) and associated with calcium network modifiers ($Q^2_{Ca}$). In the case of the Ca-containing MBG samples, the spectral assignments are less obvious because of the competing roles of $Ca^{2+}$ and $H^+$ for charge compensation of NBOs and

their associated $^{29}$Si deshielding[46,47]. $Ca^{2+}$ cations are mainly entrapped as amorphous calcium phosphates (ACP) and the $^1H \rightarrow ^{29}Si$ CP spectra showed a small signal at -83 ppm assigned to the $Q^2_{Ca}$ environment, highlighting the presence of this species close to the protons sited at the material surface. This is clearly indicative of the joint presence of $Ca^{2+}$ and $PO_4^{3-}$, resulting in ACP clusters located at the wall surface, that favoring the solubility of these materials, this is in agreement with the models previously reported[43,48]. $^{29}$Si NMR spectra 75/85-MBG-NH$_2$ display two new resonances around -60 and -68 ppm. These resonances represent silicon atoms in positions [R-*Si*(SiO)$_2$(NBO)] and [R-*Si*(SiO)$_3$], denoted as $T^2$ and $T^3$, respectively, evidencing the covalent functionalization with APTES. These signals are emphasized in the CP spectra pointing out that the linker is mainly located at the MBG surfaces. Similar $^{29}$Si NMR patterns were observed for 75/85-MBG-Lys samples. However, when the silica matrices of 75/85-MBG-Lys samples were studied ($Q^n$ signals) the relative peaks areas changed considerably compared to their immediate predecessors (75/85-MBG-NH$_2$ samples). This behaviour could be attributed to the high loss of calcium and phosphorus during the GA activation and Lys incorporation. These synthesis steps taken place in aqueous medium at room temperature overnight and in these conditions, the MBGs are very soluble[49]. The XRF results (Table 2) showed higher calcium and phosphorus decreased when the Lys was incorporated in 75/85-MBG-Lys materials. However, when the APTES was incorporated to MBG surfaces (first step of synthesis) there were hardly differences between the molar composition of 75/85-MBG and 75/85-MBG-NH$_2$ samples (Table 2). This behaviour could be due to synthesis conditions: overnight reflux of dry toluene. So, the small amount of CaO presents in these 75/85-MBG-Lys compositions could be the reason to does not exhibit $Q^2_{Ca}$ environments when the spectra are recorded by $^1$H→$^{29}$Si CP MAS NMR for 75/85-MBG-Lys samples synthetized under aqueous conditions.

$^{31}$P MAS NMR spectroscopy was used to evaluate the local environment of P atoms, thus elucidating the phosphate species that is contained in the samples. Figure S3 (Supporting Information) shows the solid-stage $^{31}$P MAS NMR spectra for 85-MBG composition (similar results were obtained for 75-MBG composition). It must be commented that CaO is a network modifier that disrupts the $SiO_2$ or $SiO_2$-$P_2O_5$ network. $Ca^{2+}$ may be dispersed into the network or forming segregated ACP clusters at the surface. All samples showed the same $^{31}$P MAS NMR patterns, a mean maximum of ~2 ppm assigned to the $q^0$ environment. This signal is typical of an ACP, where the CaO presence in these MBG matrices leads to the nucleation of an amorphous calcium phosphate[43,48]. A second weak signal sited around -7 ppm appears for these samples. This resonance falls in the range of $q^1$ tetrahedra[46-48] and can be assigned to *P*-O-P or *P*-O-Si environments. Due to the reaction conditions used in the Lys incorporation, the solubility of MBG matrices was favored, decreasing $q^0$ intensity. Thus, the phosphorus that forms the network ($q^1$) emphasized its presence in the 75/85-MBG-Lys samples.

Once the chemical nature of powdered samples was studied, their surface charge as a function of pH was evaluated by recording ζ-potential measurements at different pH values. The resulting ζ-potential *vs*. pH plots are displayed in Figure 5 and the isoelectric points (IEPs) are summarized in Table 2. The IEP is strongly related to the zero surface charge point, therefore the IEPs of 75/85-MBG, 75/85-MBG-NH$_2$ and 75/85-MBG-Lys samples will depend on their surface functionalization. 75/85-MBG samples do not have IEP. Both were charged negatively in all pHs due to the presence of silanol groups in their surface. Silanol groups, which exhibit weak acidic Brönsted character, would be deprotonated Si-OH → SiO$^-$ + H$_3$O$^+$, providing a negative surface charge. For surface modified 75/85-MBG-NH$_2$ and 75/85-MBG-Lys materials, there were greater ζ-potential variability as a function of the pH. The aminated 75/85-MBG-NH$_2$ samples presented IEPs values of 7.5 ± 0.3 and 7.8 ± 0.5 which could be explained by the presence of amine and silanol groups in the surface of samples. Three zones (A, B or C) can be observed depending on the predominate species Si-OH/Si-O$^-$ or NH$_2$/NH$_3^+$ as a function of pH range. When the amine groups were modified by an incorporation of new specie with intrinsic zwitterionic nature (Lys is an amino acid with zwitterionic nature) similar patterns were observed. For 75/85-MBG-Lys samples, the simultaneous presence of adjacent basic amine (-NH$_3^+$) and carboxylic acid groups (-COO$^-$), generated a slight change of IEPs 7.3 ± 0.1 and 7.4 ± 0.4, respectively. From ζ-potential results it can be concluded that amino functionalized (75/85-MBG-NH$_2$) and Lys incorporated materials (75/85-MBG-Lys) exhibit zero surface charges in aqueous media in the pH range of biological interest. Therefore, with the aim of evaluating the influence of the zwitterionic character provided of materials with different nature compared to non-zwitterionic 75/85-MBG materials an *in vitro* study of bacterial adhesion and cytocompatibility evaluation was carried out with *S. aureus* and MC3T3-E1 preosteoblast cells.

### 3.2 Bacterial adhesion assays

The anti-adhesive bacterial properties of non-zwitterionic and zwitterionic MBGs were investigated by carrying out *in vitro S. aureus* adhesion assays using powders pressed into disks. With the aim of normalizing the number of CFUs per surface unit, the microstructure of disk-shaped samples was investigated. Hg intrusion porosimetry measurements reveal the presence of intergranular pores with sizes in the range of 100-600 μm for 75/85-MBG samples. In case of 75/85-MBG-NH$_2$ and 75/85-MBG-Lys samples three distribution of intergranular pore were observe at 100-600 µm, 1-10 µm and 0.1-1µm range sizes for all of them. The total surface area within such ranges of macroporosity was 10, 60 and 270 cm$^2$/g for 85-MBG, 85-MBG-NH$_2$ and 85-MBG- Lys, respectively. In case of 75-MBG, 75-MBG-NH$_2$ and 75-MBG-Lys was 20, 20 and

110 cm$^2$/g, respectively. The higher porosity is attributable to the samples with organic moieties during the synthetic procedure. Therefore, the degrees of *S. aureus* adhesion for the different material surfaces, expressed as CFUs per cm$^2$, are displayed as histograms in Figure 6. Statistic one-way ANOVA tests show that the bacterial adhesion values for the different surfaces were significantly different ($p < 0.05$ and $p < 0.005$). The mechanism of bacterial adhesion inhibition on zwitterionic surfaces is not fully understood but the great polarity of the charged functional groups would enhance surface hydration via electrostatic interactions in addition to hydrogen bonding[23]. Thus, water molecules on the zwitterionic surface would create a strong repulsive force on the bacteria as they approach the surface. Figure 6 reveals that comparing 75-MBG and 85-MBG samples the different bacterial adhesion could be understood through their different composition. Bioactive glasses have some antimicrobial activity in aqueous solutions via release of their ionic compounds over time[50]. The bioactivity mechanism which consist of the release of calcium ions and the incorporation of H$_3$O$^+$ protons results in a high pH environment in closed systems which is not well tolerated by microbiota[51]. Considering 75-MBG sample exhibited higher solubility than 85-MBG, and its solubility can be attribute to double calcium content and lower network connectivity respect to 85-MBG sample, the antibacterial effect of unmodified 75-MBG sample was higher than bare 85-MBG material (Tables 2 and 3). Figure 6 also shows that the bacterial adhesion decreased 52% ($p < 0.05$) and 99.9% ($p < 0.005$) for 85-MBG-NH$_2$ and 85-MBG-Lys samples, respectively compared to bare 85-MBG material.

The combination of glass composition and zwitterionic pairs in their surface avoid almost all bacterial adhesion. The presence of -NH$_3^+$ and -COO$^-$ groups in 85-MBG-Lys and -NH$_3^+$ and -Si-O$^-$ groups in 85-MBG-NH$_2$ with an overall neutral charge at pH close to 7.4, as can be seen in ζ-potential measurements (Figure 5), results in a zwitterionic surface that could resist bacterial adhesion via a hydration layer bound through solvation of the charged terminal groups. Thus, water molecules above the zwitterionic surface would create a strong repulsive force on the bacteria as they approach the surface, which is in agreement with previous reports for ultralow fouling materials. However, the 85-MBG-Lys sample exhibit better low-fouling properties than 85-MBG-NH$_2$. Izquierdo-Barba *et al.*[21] demonstrated a similar behaviour in a SBA-15 zwitterionic surface that resist bacterial adhesion in presence of NH$_3^+$/-COO$^-$ pairs with significantly higher anti-adhesive bacterial properties than NH$_3^+$/-SiO$^-$ pairs. Moreover, 75-MBG-NH$_2$ sample present no differences compared to 75-MBG-Lys. The difference between 75-MBG-NH$_2$ and 85-MBG-NH$_2$ samples on their physico-chemical properties is their different calcium amount. That was enough to decrease (4.2 folds) the bacterial adhesion in 75-MBG sample compared to 85-MBG samples. 75-MBG-NH$_2$ sample, in addition to zwitterionic surface, present a faster interchange of calcium ions and protons in their surface, due to its glass composition, which could not be well tolerated by bacteria as is explained above. The second difference is the intergranular spaces on disks that are 3 folds higher in 85-MBG-NH$_2$ than in 75-MBG-NH$_2$ so the surface disposable to bacterial environment are much higher. The visualization of adhered bacteria into the different surfaces was performed by SEM. Figure 7 shows SEM micrographs of disk-shaped 85-MBG, 85-MBG-NH$_2$ and 85-MBG-Lys materials after bacterial adhesion assays. SEM micrographs reveal the beginning of bacterial colonization by *S. aureus* on the 85-MBG and 85-MBG-NH$_2$ surfaces. The presence of bacteria onto 85-MBG-Lys sample was near to zero and in addition, the bacteria found showed membrane damage. Figure 7 also reveals that these *S. aureus*, with an average size of 0.5-0.8 μm, do not penetrate into the 85-MBG, 85-MBG-NH$_2$ and 85-MBG-Lys cavities existing between the material particles, which are higher than 1 μm, as determined by Hg intrusion porosimetry. However, the non-zwitterionic 85-MBG surface seems to be much more attractive by bacterial adhesion. The critical stage of biofilm formation is the bacterial adhesion. Many aspects of bacterial adhesion mechanism have not been well understood due to the numerous physicochemical factors involved[52]. This step has been reported as a prior stage to subsequent biofilm formation, to avoid it is a critical step to avoid bone infection[37].

### 3.3 *In vitro* cytocompatibility assays

To test cytocompatibility, *in vitro* analyses were performed using MC3T3-E1 preosteoblast cells on different zwitterionic surfaces (the materials were conformed in disk-shaped pieces). Preosteoblast cells in absence of any material were performed as a control. An essential requisite for a biocompatible material is the capability of allowing cell proliferation in its proximity and cellular adhesion and proliferation on its surface. Cells in contact with an appropriate surface will firstly attach, then adhere, and finally spread[53]. Figure 8A shows the morphology of MC3T3-E1 onto disks surfaces. Cells attached and proliferated around disks samples and maintained their typical preosteoblast morphology in presence of 85-MBG, 85-MBG-NH$_2$ and 85-MBG-Lys samples after 24h. Similar results were obtained from the 75-MBG batch of samples (Figure S4, Supporting Information). Figure 8B shows cell viability in terms of mitochondrial activity after 2 and 4 days of culture. MBG materials showed an increase in the levels of proliferation as a function of incubation time on both non-zwitterionic and zwitterionic samples. Proliferation was significantly higher in aminated 85-MBG-NH$_2$ than the rest of samples (p < 0.05) after 4 days. In this case, 85-MBG-NH$_2$ sample presented a lower release of calcium to the medium respect to the 75-MBG series due to lower calcium in its structure. M. Alcaide *et al.*[54] found that for MBG materials a supersaturated fluid, especially in Ca$^{2+}$ cations could be connected with the cytostatic effect over osteoblasts, although no cytotoxicity was detected. This effect has been observed until 3 days of assay with osteoblast cells. Moreover, Cicuéndez *et al.*[55] observed that adsorption of serum protein is much higher for the aminated surfaces compared to unmodified surfaces of MBG materials. Amine functionalized surfaces display increased adsorption of serum proteins, such as fibrinogen and fibronectin. Thus, the mechanism for promoting increased cell development of the 85-MBG-NH$_2$ aminated samples is likely facilitated by the formation of a conditioning protein layer.

To evaluate the cytotoxic effect of the different materials, the amount of LDH released to osteoblast-like cells cultured in the presence of the different scaffolds was determined after 2 and 4 days (Figure 8C). The results indicated that no cytotoxic agent was released compared to control on MC3T3-E1 preosteoblast cells. Figure 8D shows the ALP activity after 4 days of MC3T3-E1 cells incubation on MBG disks. ALP is an early marker of the osteoblast phenotype and is up regulated at the onset of osteoblast differentiation[56]. Significant differences (p < 0.05) from that of the control except on the 85-MBG-Lys surface (p < 0.05) were obtained. The results revealed that the positive effect on osteoblast differentiation for cells attached to sample surfaces containing Lys. 75-MBG-Lys and 85-MBG-Lys samples present Lys anchored in their surfaces but higher amount in 85-MBG-Lys (12.8 w%) than 75-MBG-Lys sample (9.0 w%). So, this effect can be attributed to Lys enhanced differentiation[57]. Other biomaterials with similar Lys functionalization have been reported, such as phosphoserine-tethered poly(epsilon-lysine) dendrons, when applied as a film to titanium surfaces, enhance the differentiation of osteoblastic cells and the activation of Wnt/b-catenin signaling[58]. Moreover, Fini *et al.* reported that Lys exhibits therapeutic effects observed *in vivo* in osteoporosis and fracture healing[59]. The cells adhesion and spreading onto zwitterionic and unmodified surfaces could be visualized by fluorescence microcopy (Figure 9). Actin cytoskeleton was stained with phalloidin (red) and the nuclei with DAPI (blue). Atto 565-phalloidin stains the F-actin microfilaments to visualize the cytoskeleton. DAPI is a DNA dye widely used to visualize changes in the chromatin in cells, indicating cell viability. As a consequence of the phalloidin absorption by the 75/85-MBG-Lys samples, actin filaments could not be observed (Figure 9), although nuclei stained with DAPI could be visualized. The images show viable and well-spread cells which have maintained their typical preosteoblast morphology, MC3T3-E1 cells fully colonized the non-zwitterionic and zwitterionic MBG surfaces.

It should be highlighted that the development of surfaces with simultaneous opposite responses toward cells and bacteria proliferation would represent a significant achievement in orthopedic implantology[60]. To understand the competition between eukaryotic cells and bacteria for colonizing the surface, known as "the race for the surface"[61], is a critical problem, largely unsolved up to now. The way eukaryotic cells interact with a surface is through an interface of integrins that interact with specific moeties of the extracellular matrix, such as Arginyl-glycyl-aspartic acid (RGD) motifs[62]. In this sense, the surface chemistry at the nanoscale could be a determining factor in which type of integrin will mediate cell adsorption onto the biomaterial surface[63]. On the contrary, bacteria are prokaryotic cells and their cell wall is composed of phospholipids, like eukaryotic cells, but bacteria present an external layer of peptidoglycan much more rigid. Moreover, size bacteria are much smaller than eukaryotic cells[64]. It has been demonstrated that both specific features of bacteria, might be key factors accounting for their capability to perceive variations in the chemical and topological features of biomaterial surfaces at the nanoscale[60]. In terms of surface adhesion, whereas single eukaryotic cells can adhere to surfaces independently of each other through integrins, bacteria live on surfaces as a community within a resistant-extracellular polymeric substance, named as bacteria biofilm. Bacterial adhesion, the first step of biofilm formation, is governed mainly by the electrostatic attraction forces between bacteria and surfaces (mediated by adhesins). As a consequence the electrochemical nature of the biomaterials plays a major role[65,66]. Therefore, zwitterionic surfaces are capable of forming a hydration layer that would act as barrier to hinder bacterial adhesion, which would allow preventing infection. In fact, recently, opposite behaviour against bacterial and osteoblast cells in different biomaterials (SBA-15 and hydroxyapatite) with zwitterionic nature has been described[7,21] and also in this research work extrapolated to MBG materials. Finally, the results obtained reveal that after cellular surface colonization such non-zwitterionic and zwitterionic MBGs have the capability to allow preosteoblast cells development ensuring the functionality of these systems for bone tissue regeneration purposes.

## 4. Conclusions

Zwitterionization constitutes an incipient tool for the design of advanced bioceramics for biomedical applications. This manuscript reports for the first time the design and synthesis of zwitterionic Mesoporous Bioactive Glasses (MBGs) with bacterial anti-adhesive properties and *in vitro* cytocompatibility behaviour. MBGs surfaces have been prepared by post-synthesis bifunctionalization of a ternary $SiO_2$-$CaO$-$P_2O_5$ system with an aminated APTES followed by an amino acid (Lys) anchored. The proper design of the functionalization strategy allows $-NH_3^+$/$-SiO^-$ zwitterionic pairs and $-NH_3^+$/$-COO^-$ zwitterionic pairs for 75/85-MBG-$NH_2$ and 75/85-MBG-Lys, respectively with zero point charge at physiological pH. It has been demonstrated that the efficient interaction of these zwitterionic pairs onto MBGs surfaces reduces *S. aureus* adhesion by 99.9% in 75-MBG-NH$_2$, 75-MBG-Lys and 85-MBG-Lys glasses. *In vitro* cytocompatibility of zwitterionic and non-zwitterionic MBGs is preserved after being submitted to the functionalization process, as demonstrated by cell culture with MC3T3-E1 preosteoblast cells. This research work highlights the versatility of the functionalization method developed here to prepare zwitterionic MBGs, which could be applied to materials available in different shapes, such as powders, particles, granules, coatings, dense blocks, or even 3D scaffolds, depending on the targeted clinical application. We envision that the "zwitterionization" of MBGs could represent an added value among the current properties of MBGs, since the decrease of *S. aureus* adherence would lead to a decrease in the infection rates after implantation surgery.

## 5.- Acknowledgements

MVR acknowledges funding from the European Research Council (Advanced Grant VERDI; ERC-2015-AdG Proposal No. 694160). The authors also thank to Spanish MINECO (MAT2015-64831-R) and Instituto de Salud Carlos III (PI15/00978). The authors wish to thank the ICTS Centro Nacional de Microscopia Electrónica (Spain), CAI X-ray Diffraction and CAI nuclear magnetic resonance of the UCM (Spain) for the assistance.

**Scheme 1.** Reactions sheme of first (aminated) and second functionalization (zwitterionization) of 75-MBG and 85-MBG samples.

**Figure 1.** (A) XRD patterns of non-zwitterionic and zwitterionic 75/85-MBG samples. (B) TEM images of wormlike mesoporous sample 75-MBG-Lys and TEM images acquired with the electron beam perpendicular and parallel (inset) to the mesoporous channels for 85-MBG zwitterionic sample.

**Figure 2.** $N_2$ adsorption-desorption isotherms and the corresponding pore size distributions of the different materials.

**Figure 3.** $^1H \rightarrow ^{13}C$ CP/MAS solid state NMR spectra of 75/85-MBG-$NH_2$ and 75/85-MBG-Lys samples.

**Figure 4.** Solid-state $^{29}Si$ single-pulse (left) and cross-polarization (right) MAS NMR spectra (with their $Q^n$ and $T^n$ silicon environments shown at the top) corresponding to the 75-MBG and 85-MBG samples before and after its surface functionalization. The areas for the $Q^n$ and $T^n$ units were calculated by Gaussian line-shape deconvolutions.

**Figure 5.** ζ-Potential *vs* pH plots of the different 75-MBGs and 85-MBGs. A, B and C indicate the amino, silanol and carboxylic species of MBG-$NH_2$ and MBG-Lys from 75-MBG and 85-MBG samples at different pHs.

**Figure 6.** Schematic draw of low-fouling zwitterionic effect. *S. aureus* attachment to the 75-MBG and 85-MBG non-zwitterionic and zwitterionic samples surfaces. Bacterial assay are expressed as means standard deviations of the three independent experiments indicated in each case. Statistical analysis was performed using the Statistical Package for the Social Sciences (SPSS) version 22 sofware (IBM). Statistical comparisons were made by analysis of variance (ANOVA). The Scheffé test was used for post hoc evaluation of differences between groups. In all statistical evaluations, $p < 0.05$ (*) and $p < 0.005$ (***) were considered as statistically significant.

**Figure 7.** SEM micrographs of the morphology of bacteria adhered to the sample surfaces.

**Figure 8.** Biocompatibility cell culture tests. (A) Cell morphology evaluation by optical microscopy of MC3T3-E1 cells cultured after 1 day. Asterisks indicate significant differences between the significative diferent samples. Comparisons between control and zwiterionic MBGs. Statistical significance: $p < 0.05$ (*). (B) MTS assays of preosteoblast cells after 2 and 4 days of culture under standard conditions. (C) LDH released into the medium from MC3T3-E1 cells cultured. (D) Cellular differentiation assays measuring ALP activity (ALP U/L) after 4 days under standard conditions. The values shown are means standard errors for all the assays. * symbol indicates significant differences between control and MBG samples. º symbol corresponds to significant differences between 85-MBG-Lys and rest of samples ($p < 0.05$).

**Figure 9.** Morphology evaluation by confocal microscopy of MC3T3-E1 preosteoblasts cultured onto 75/85-MBG, 75/85-MBG-NH$_2$ and 75/85-MBG-Lys mesoporous glasses. Cells were stained with DAPI (blue) for the visualization of the cell nuclei and phalloidin-Atto 488 (red) for the visualization of cytoplasmic F-actin filaments.

**Table(1)**

**Table 1.** Amounts of reactants used for the synthesis of MBGs via EISA method. Nominal (% mol) and XRF-analyzed compositions of the MBGs.

| Sample | Precursors[a] | | | Nominal composition (% mol) | Composition (XRF)[b] | | |
|---|---|---|---|---|---|---|---|
| | TEOS (mL) | TEP (mL) | $Ca(NO_3)_2 \cdot 4H_2O$ (g) | | $SiO_2$ (% wt) | CaO (% wt) | $P_2O_5$ (% wt) |
| 75-MBG | 14.38 | 1.57 | 4.05 | 75 $SiO_2$ – 20 CaO – 5 $P_2O_5$ | 75.6 ± 2.8 | 18.5 ± 2.5 | 5.9 ± 0.3 |
| 85-MBG | 15.85 | 1.27 | 1.96 | 85 $SiO_2$ – 10 CaO – 5 $P_2O_5$ | 85.5 ± 0.9 | 10.4 ± 0.8 | 4.1 ± 0.2 |

[a] All the syntheses were carried out with 8 g of P123, 2 mL of HCl 0.5 N, and 120 mL of absolute ethanol.

[b] The percentages in weight of $SiO_2$, CaO and $P_2O_5$ of the resulting materials after being calcined were determined by XRF measurements.

**Table 2.** Textural properties derived from $N_2$ sorption measurements for bare MBG materials and the differently surface functionalized samples[a]. Percentage of surface functionalization, XRF measurements and isoelectric point of samples[b].

| Sample | $S_{BET}$ (m²/g) | $V_T$ (cm³/g) | $D_P$ (nm) | $d_{100}$ (nm) | $a_0$ (nm) | $t_{wall}$ (nm) | % funct. TG | Composition (XRF) | | | IEP |
|---|---|---|---|---|---|---|---|---|---|---|---|
| | | | | | | | | $SiO_2$ (% wt) | CaO (% wt) | $P_2O_5$ (% wt) | |
| **75-MBG** | 375 | 0.53 | 5.6 | 7.9 | 9.1 | 3.5 | -- | 75.6 | 18.5 | 5.9 | -- |
| **75-MBG-NH₂** | 235 | 0.38 | 5.8 | 8.2 | 9.5 | 3.7 | 5.6 | 78.3 | 16.1 | 5.6 | 7.5 ± 0.3 |
| **75-MBG-Lys** | 255 | 0.31 | 4.7 | 8.1 | 9.3 | 4.6 | 9.0 | 95.3 | 1.9 | 2.8 | 7.3 ± 0.1 |
| **85-MBG** | 415 | 0.53 | 5.2 | 7.2 | 8.3 | 3.1 | -- | 85.5 | 10.4 | 4.1 | -- |
| **85-MBG-NH₂** | 275 | 0.36 | 4.7 | 7.5 | 8.7 | 4.0 | 4.0 | 85.4 | 10.5 | 4.1 | 7.8 ± 0.5 |
| **85-MBG-Lys** | 320 | 0.31 | 3.8 | 7.4 | 8.5 | 4.7 | 12.8 | 98.5 | 0.8 | 0.7 | 7.4 ± 0.4 |

[a] $S_{BET}$ is the total surface area determined by the BET method between the relative pressures (P/P⁰) 0.05-0.25. $V_T$ is the total pore volume obtained using the t-plot method. $D_P$ is the pore diameter calculated by means of the Barrett-Joyner-Halenda (BJH) method from the adsorption branch of $N_2$ isotherm. $d_{100}$ is the $d$-spacing of the 100 diffraction. $a_0$ is the unit cell parameter calculated by XRD, being $a_0 = 2 \cdot d_{100}/\sqrt{3}$ and $t_{wall}$ is the wall thickness calculated using the equation $t_{wall} = a_0 - D_P$ for hexagonal *p6mm* structures.

[b] The percentage of functionalization with APTES and later Lys incorporation was estimated from thermogravimetric (TG) measurements. The percentages in weight of $SiO_2$, CaO and $P_2O_5$ of samples were determined by XRF measurements. IEP point is the isoelectric point of samples were determined by ζ-potential measurements.

Table(3)**Table 3.** Chemical shifts and relative peak areas obtained by solid state $^{29}$Si Single Pulse MAS NMR spectroscopy. The peaks areas for the $Q^n$ and $T^n$ units were calculated by Gaussian line-shape deconvolutions and their relative populations were expressed as percentages (%).

| Sample | $^{29}$Si Single Pulse MAS NMR | | | | | $<Q^n>^c$ | $C^d$ |
|---|---|---|---|---|---|---|---|
| | $T^2$ (%) | $T^3$ (%) | $Q^2$ (%) | $Q^3$ (%) | $Q^4$ (%) | | |
| **75-MBG** | -- | -- | -91.3 (6.1) | -101.9 (39.9) | -111.1 (54.0) | 3.48 | -- |
| **75-MBG-NH$_2$** | -60.1 (1.7) (22.2)$^a$ | -68.1 (5.8) (77.8)$^a$ | -91.2 (9.2) (10.0)$^b$ | -101.4 (39.6) (43.8)$^b$ | -111.7 (43.7) (47.2)$^b$ | 3.40 | 92.6 |
| **75-MBG-Lys** | -60.7 (2.0) (43.8)$^a$ | -66.9 (2.6) (56.2)$^a$ | -93.3 (5.9) (6.2)$^b$ | -102.0 (33.5) (35.2)$^b$ | -112.0 (56.0) (58.6)$^b$ | 3.52 | 85.4 |
| **85-MBG** | -- | -- | -93.0 (3.4) | -101.3 (19.0) | -111.0 (77.6) | 3.74 | -- |
| **85-MBG-NH$_2$** | -59.1 (5.0) (46.6)$^a$ | -67.5 (5.7) (53.4)$^a$ | -95.0 (10.3) (11.5)$^b$ | -101.7 (22.0) (24.7)$^b$ | -111.6 (57.0) (63.8)$^b$ | 3.52 | 84.5 |
| **85-MBG-Lys** | -55.4 (4.3) (60.6)$^a$ | -64.5 (2.8) (39.4)$^a$ | -91.6 (4.2) (4.5)$^b$ | -102.8 (40.9) (44.0)$^b$ | -112.6 (47.8) (51.5)$^b$ | 3.47 | 79.8 |

$^a$ indicate the % amount of only T species present in the functionalized surface agent.
$^b$ indicate the % amount of only Q species present in silica MBG networks.
$^c$ network connectivity of MBGs as a function of chemical composition $<Q^n> = (4\times\%Q^4)/100 + (3\times\%Q^3)/100 + (2\times\%Q^2)/100 + (\%Q^1)/100$
$^d$ condensation degree based on $^{29}$Si MAS NMR-Single Pulse data $C = 1/3[(\%T^1)+(2\times\%T^2)+(3\times\%T^3)]$

**Scheme 1**

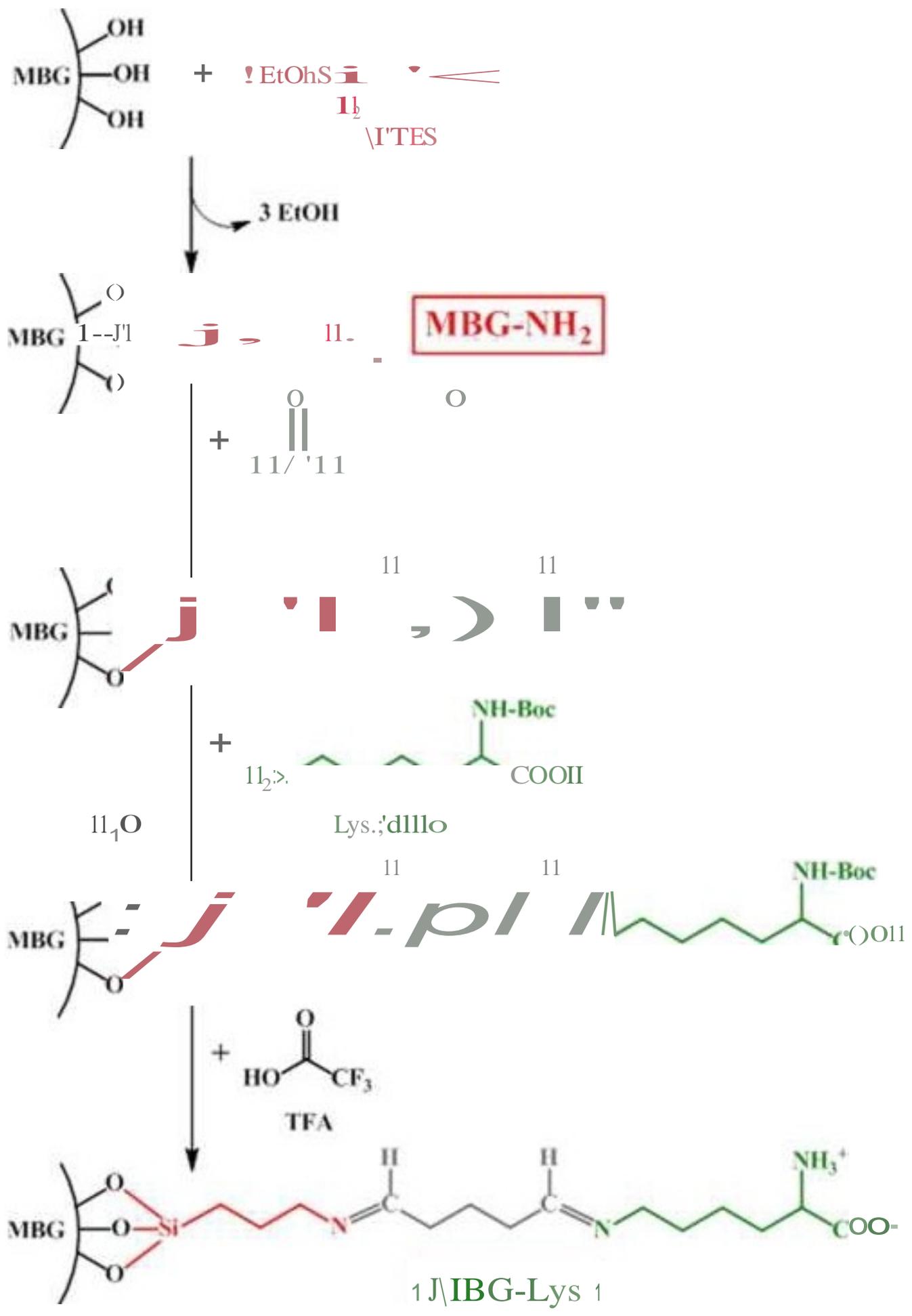

(A)
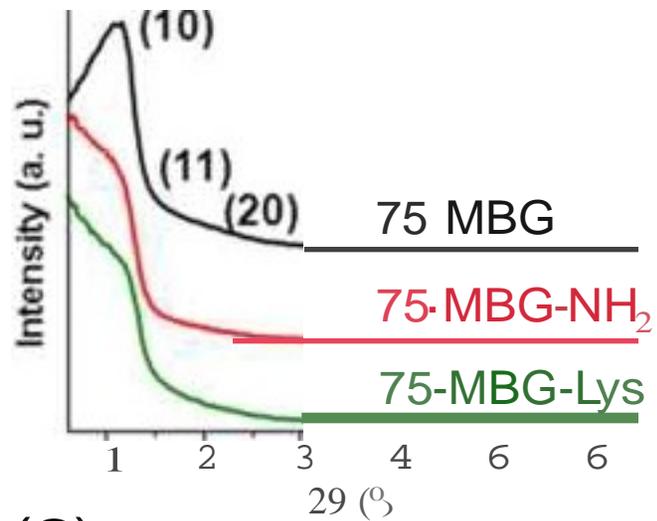
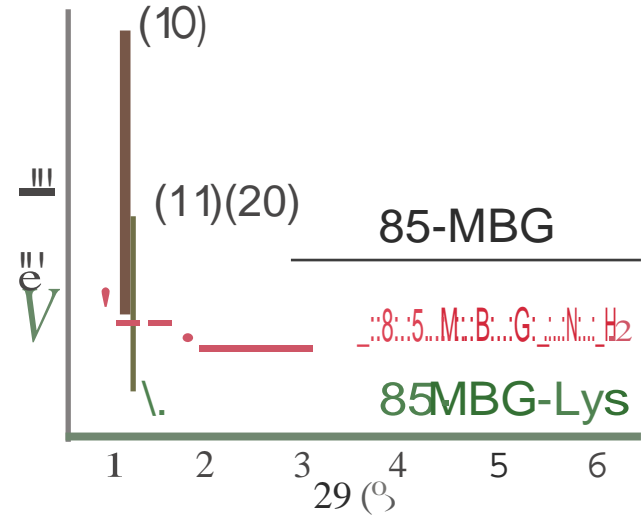

(B)
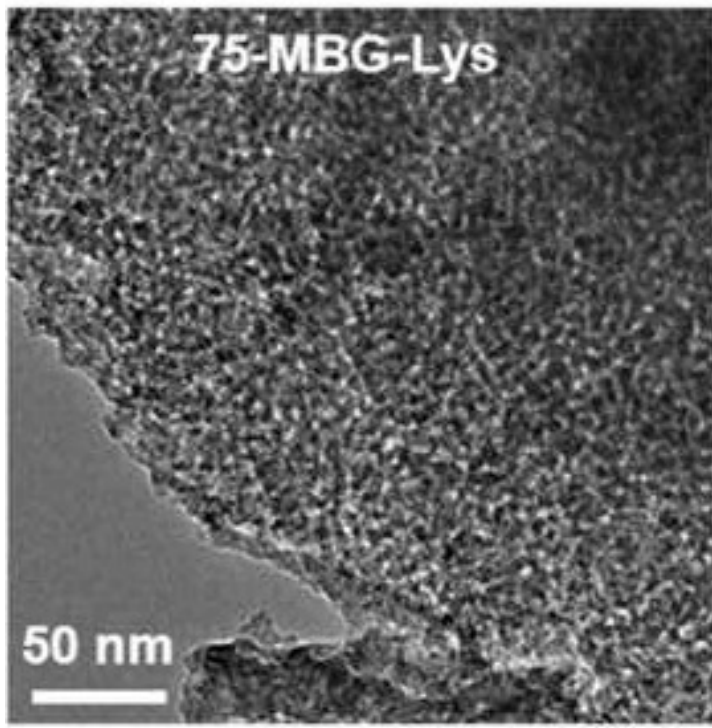
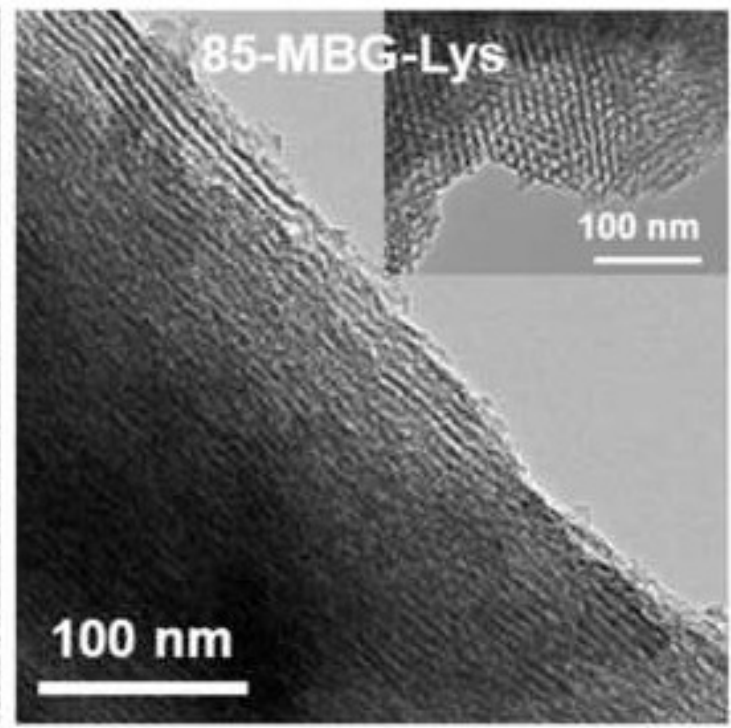

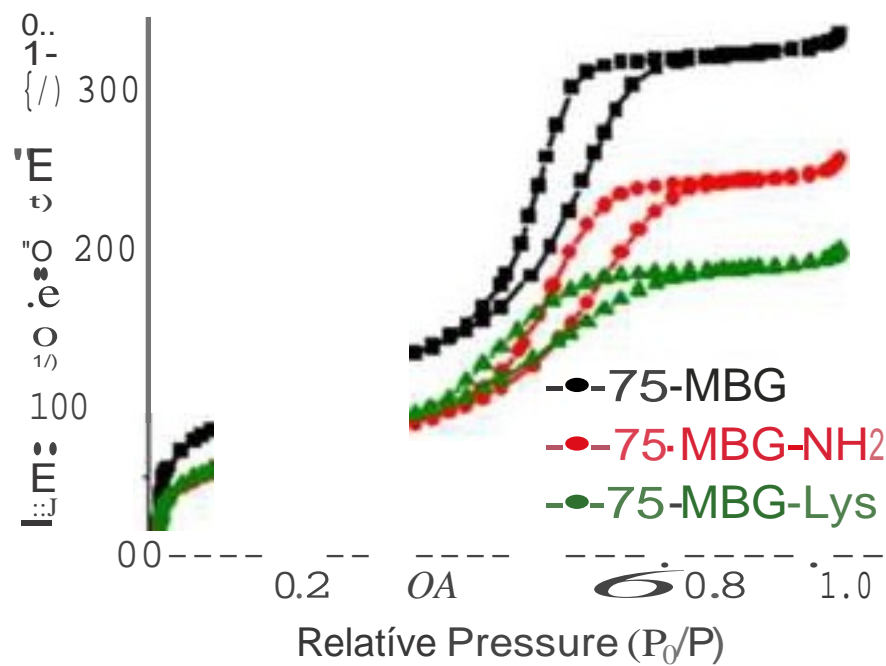
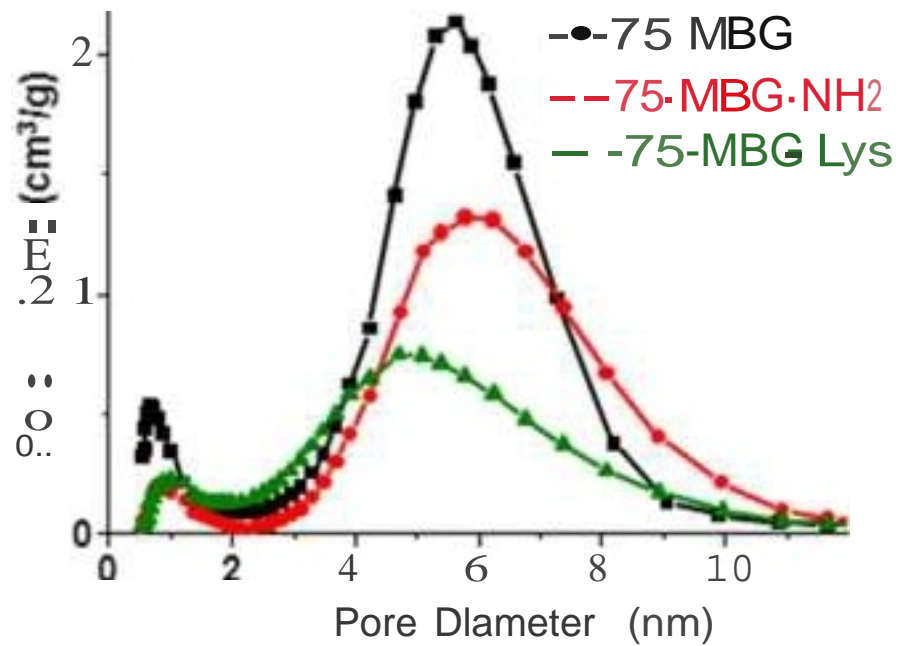
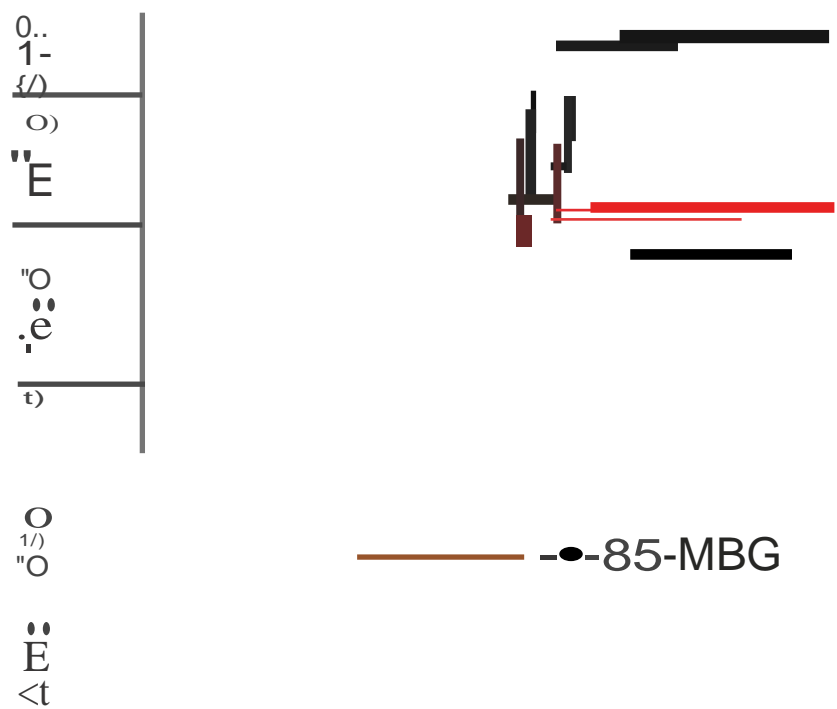
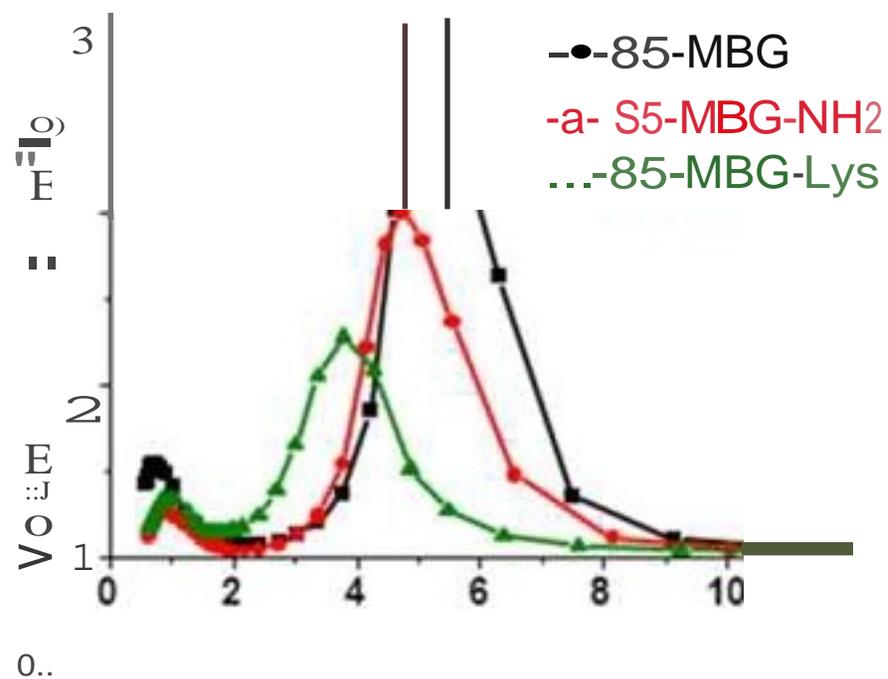

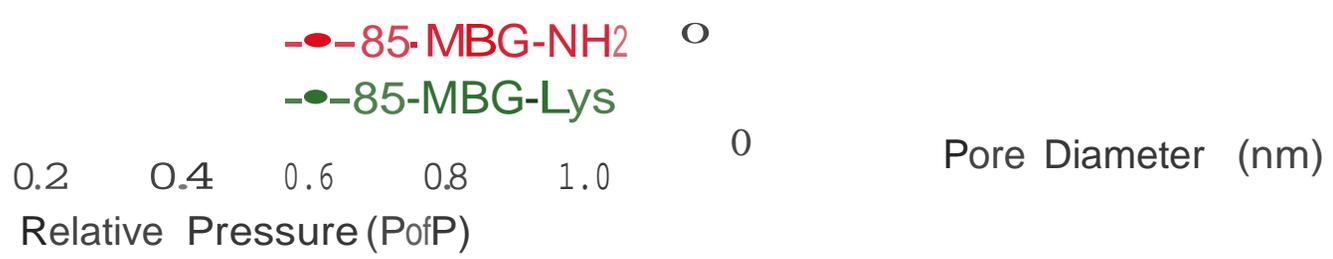

-●- 85-MBG-NH2
-●- 85-MBG-Lys

0.2  0.4  0.6  0.8  1.0
Relative Pressure (P/P₀)

Pore Diameter (nm)

**Figure3**

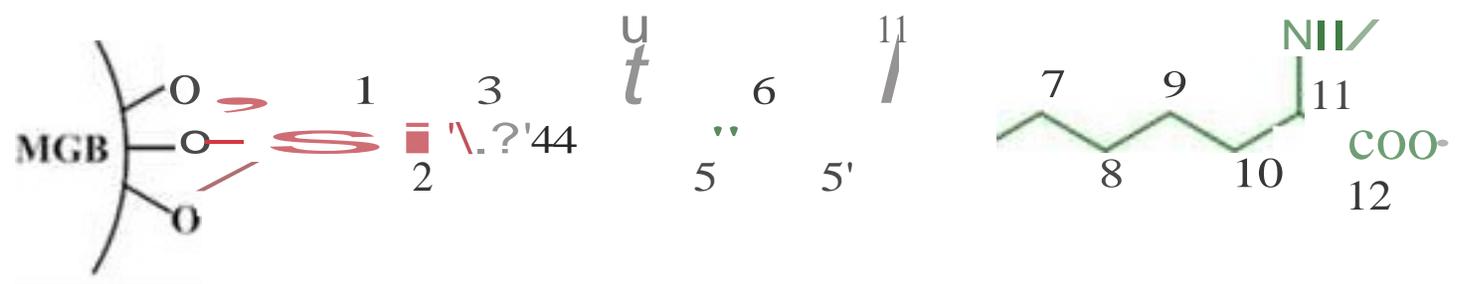
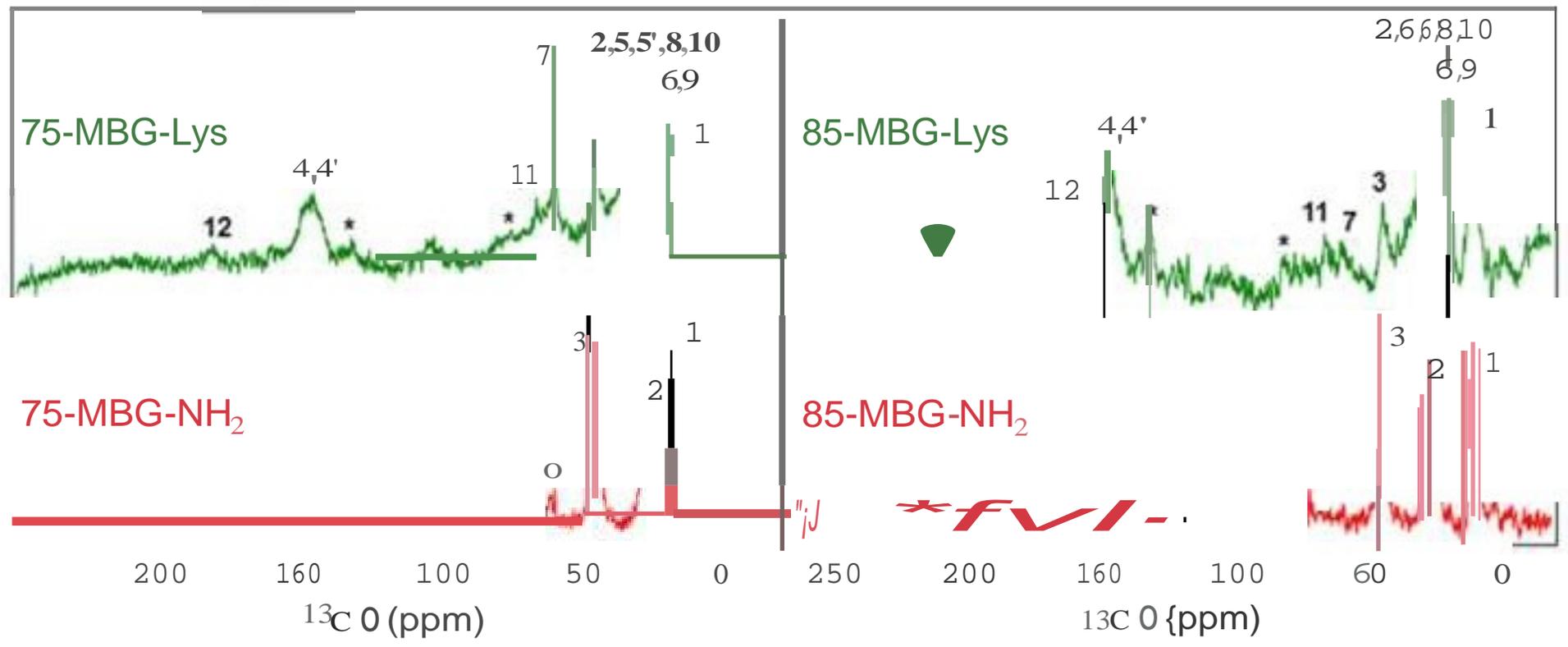

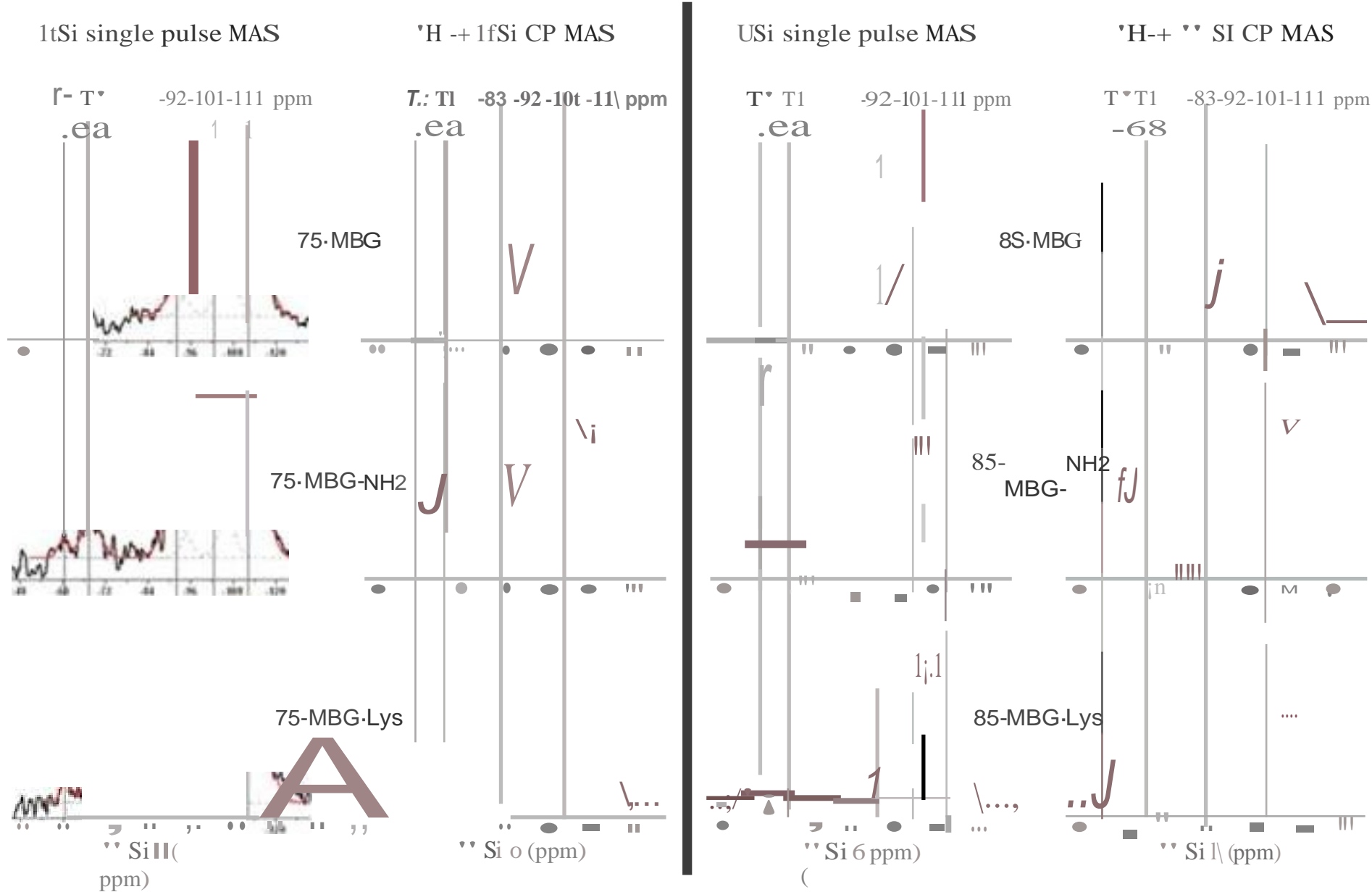

Figure 5

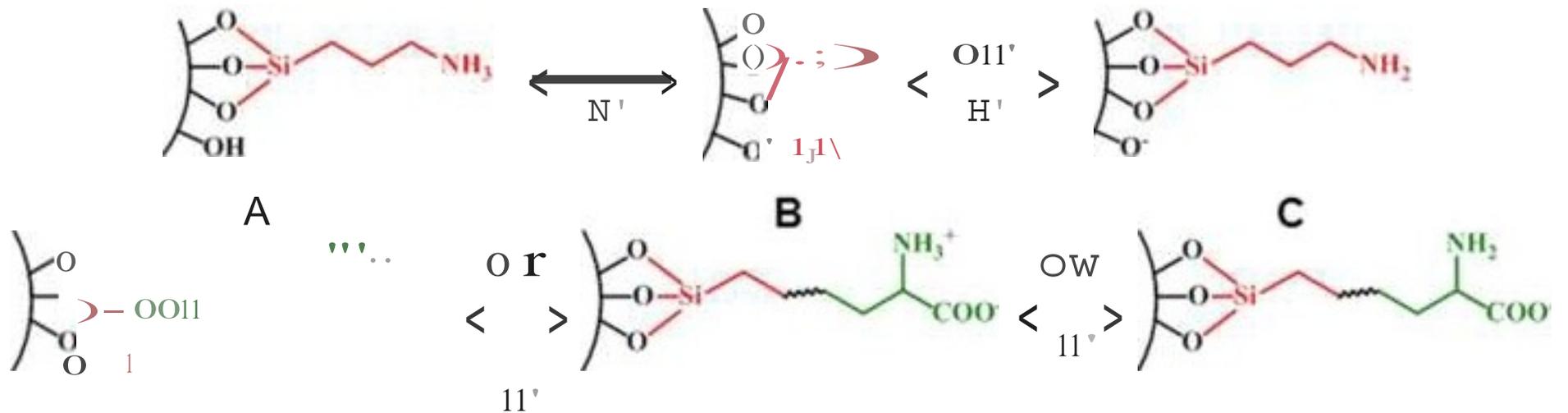
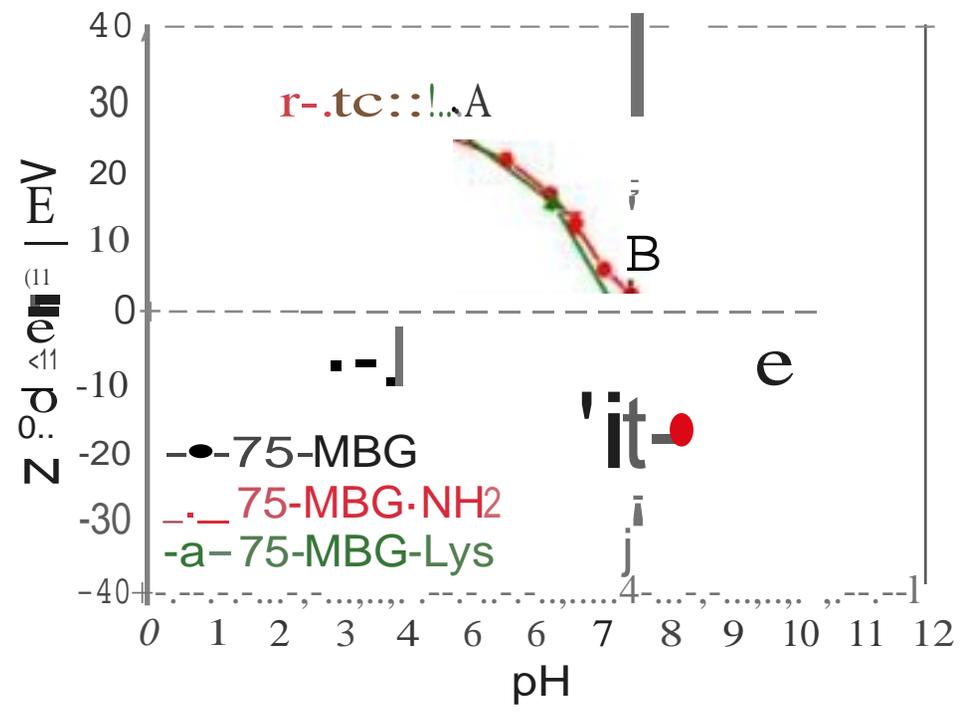
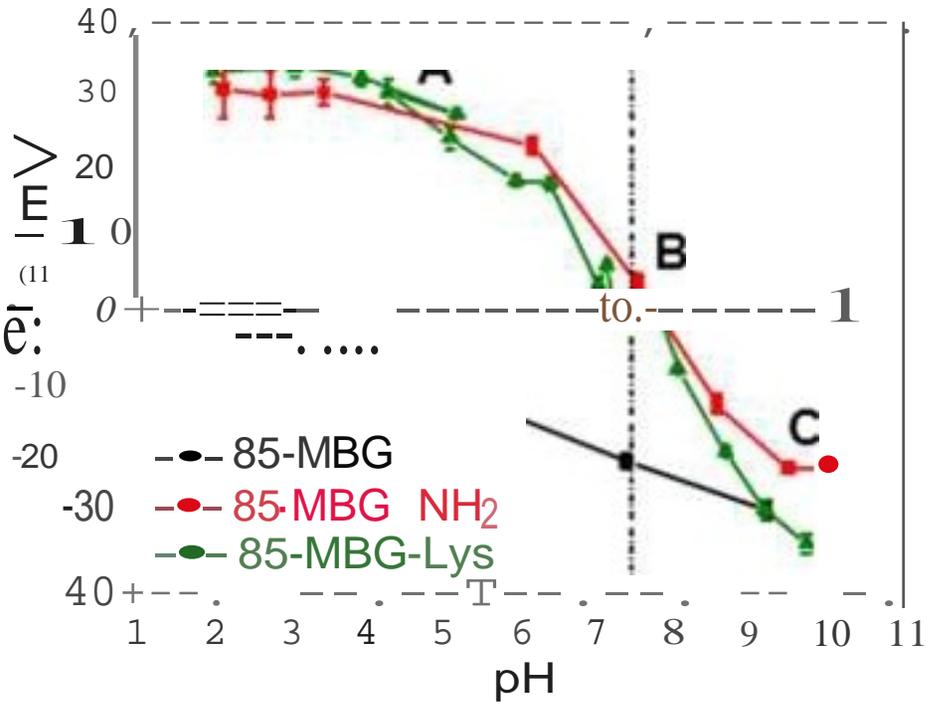

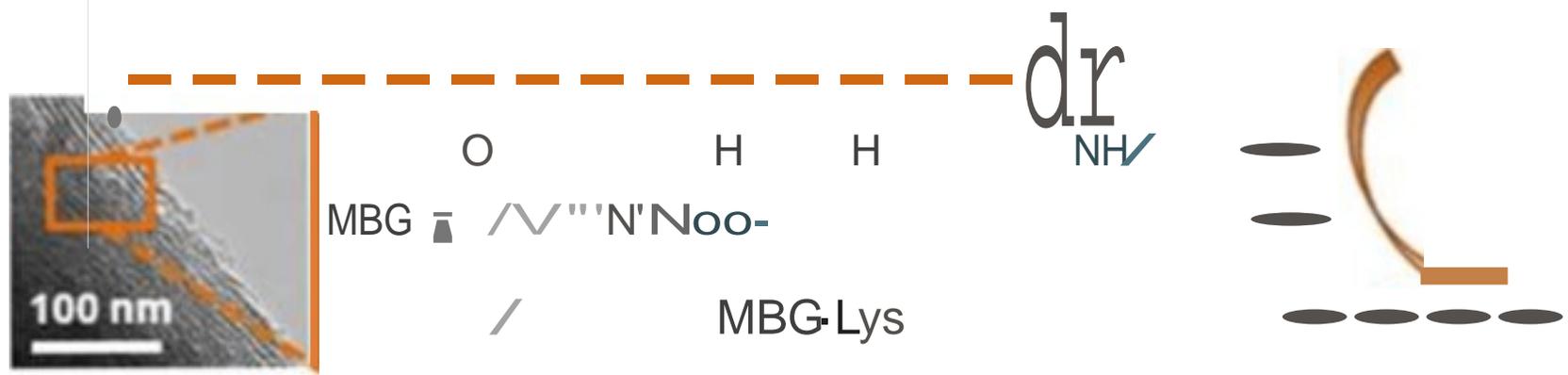
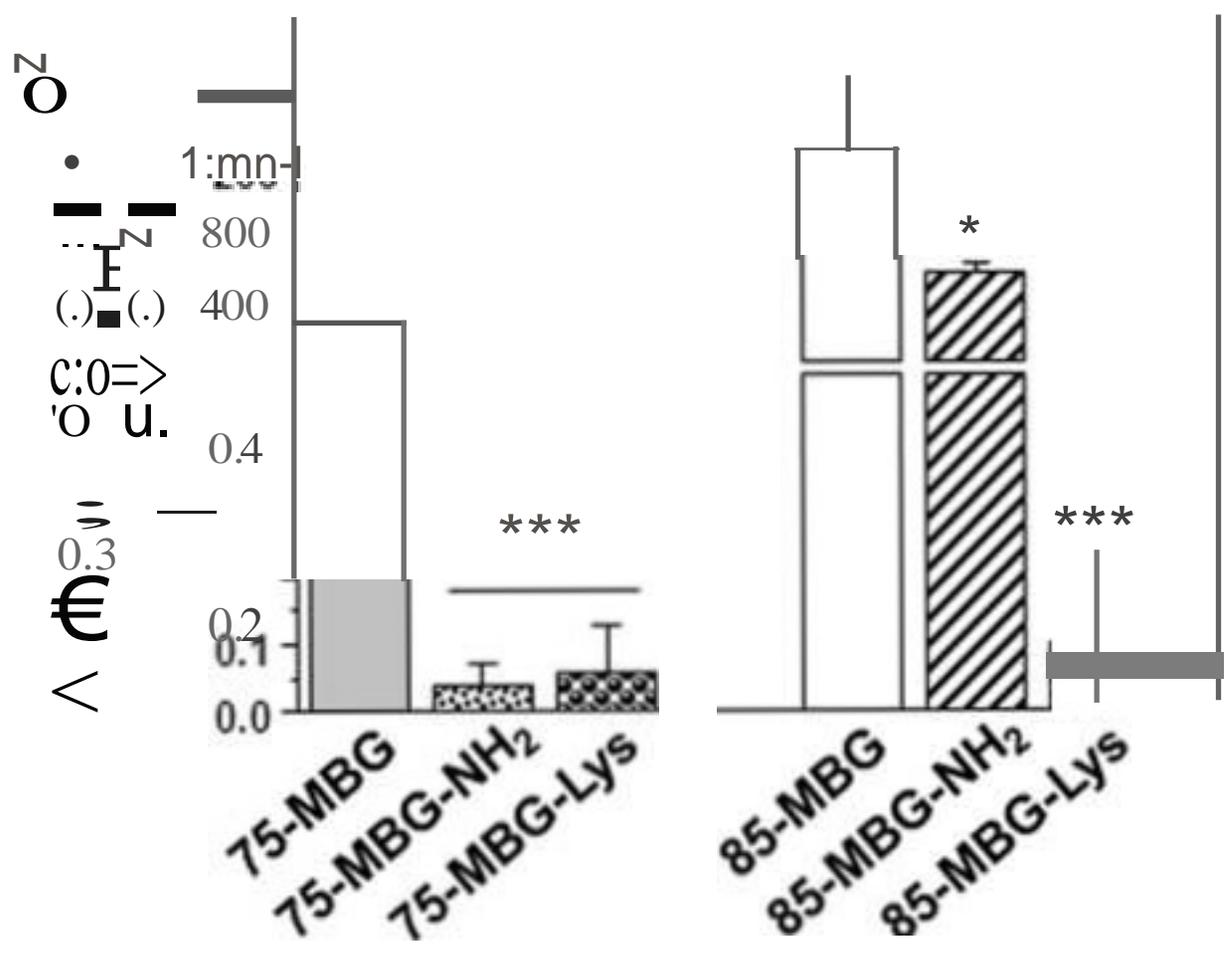

Figure 7
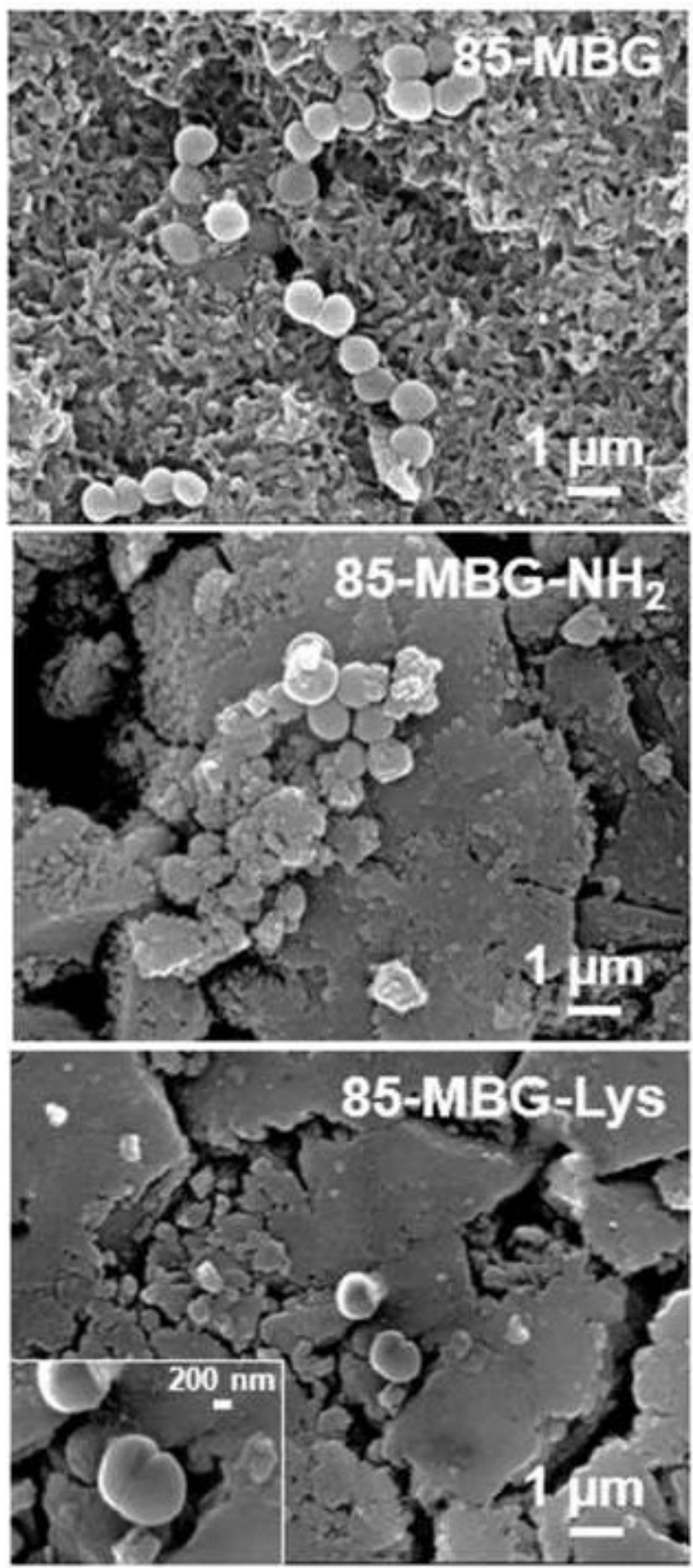

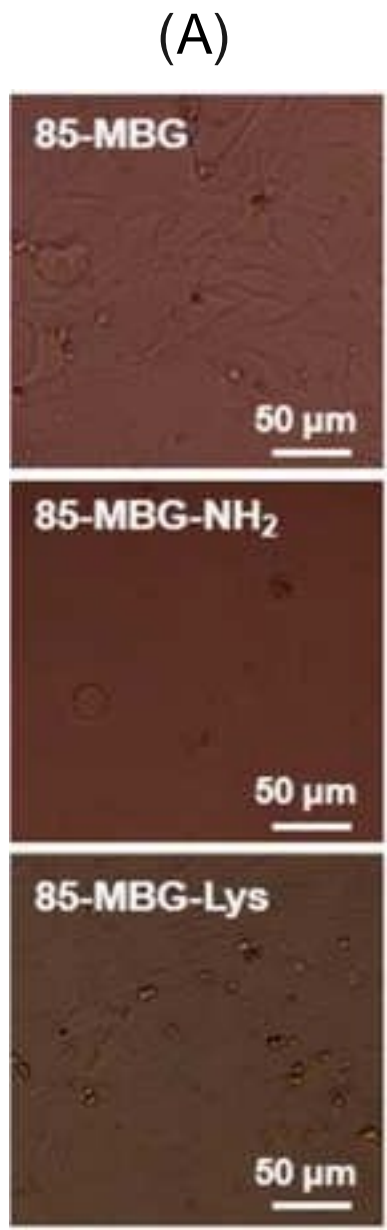
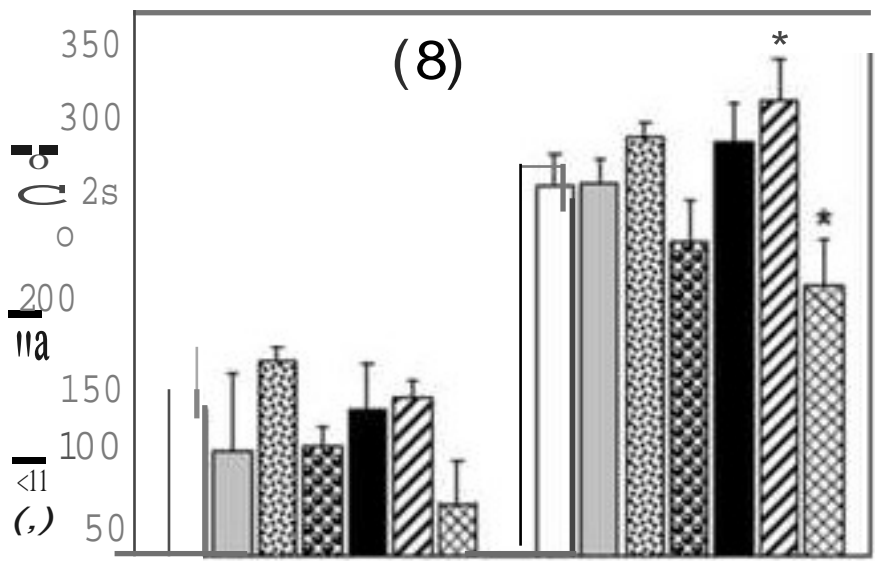
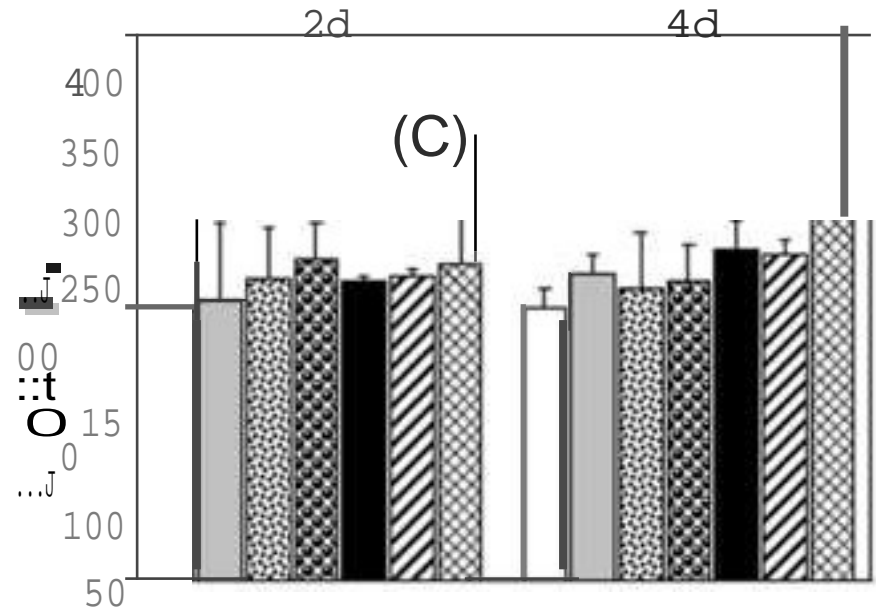
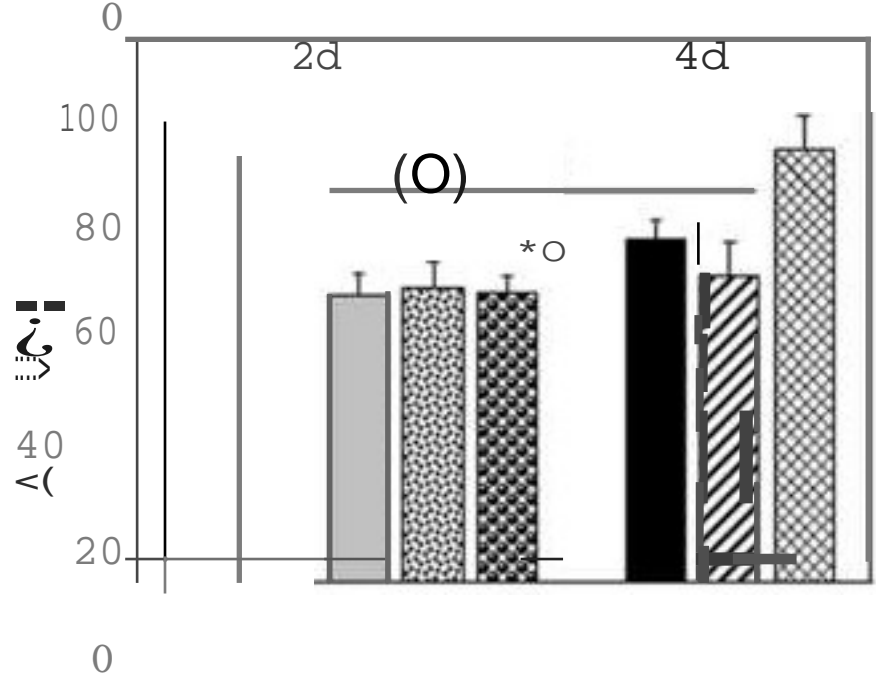

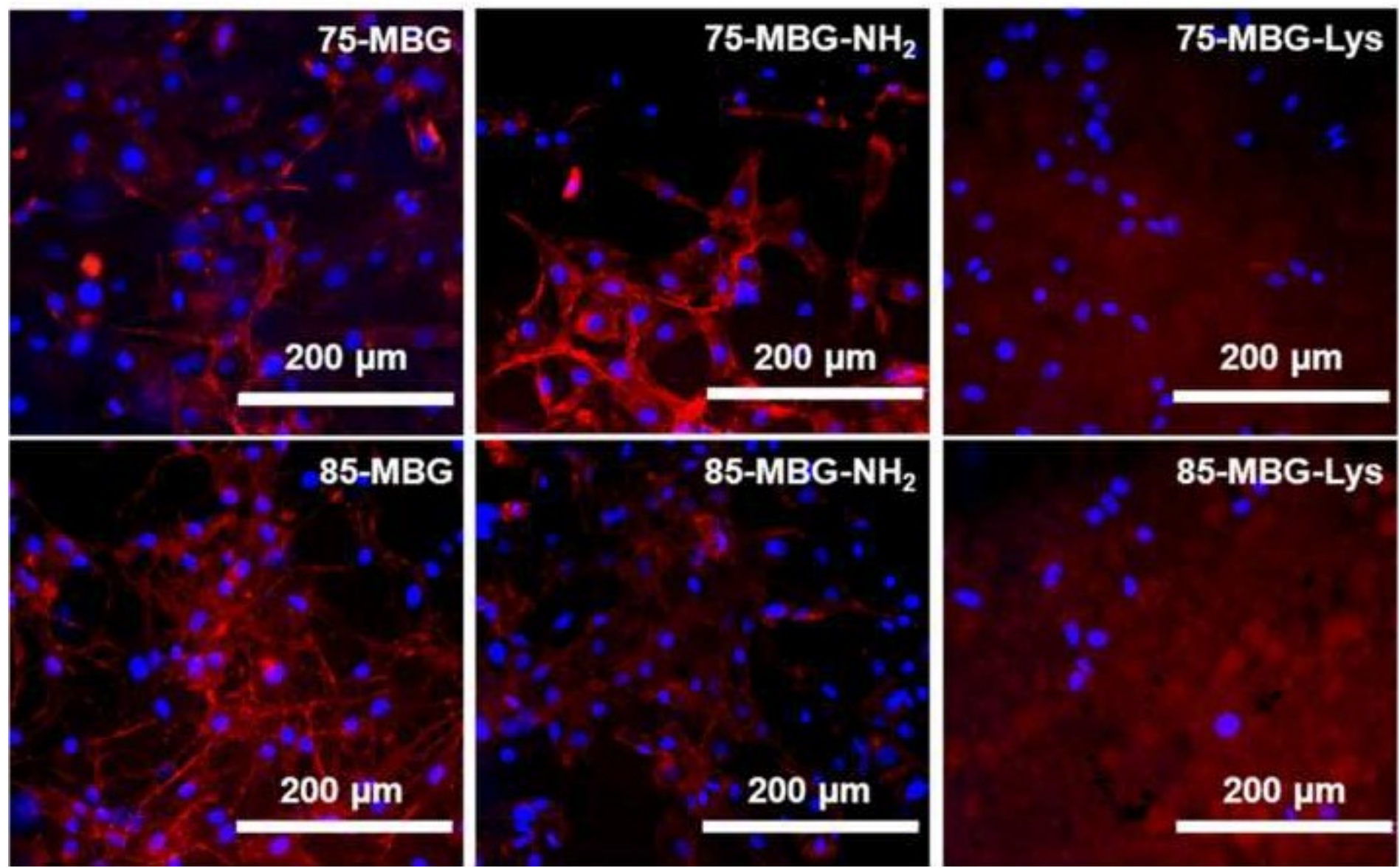

**Supplementary Material**

[Click here to download Supplementary Material: AB-17-355 Supporting Information.docx](#)

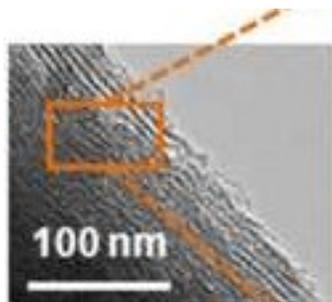 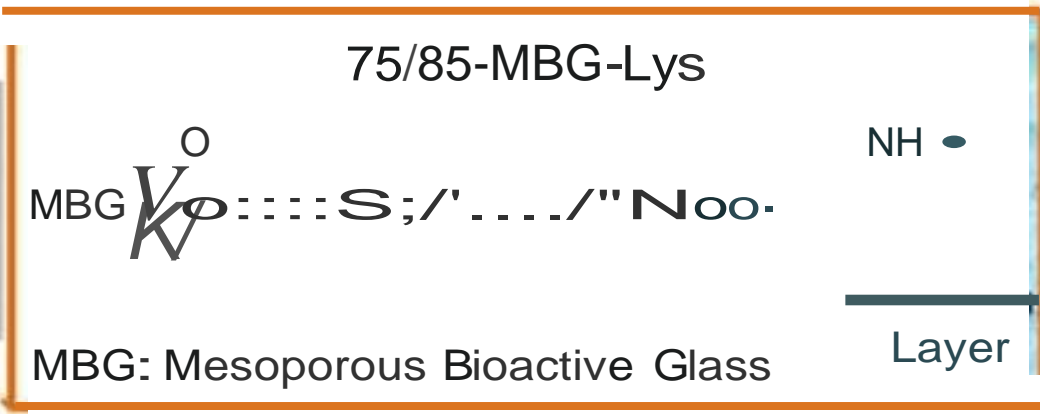 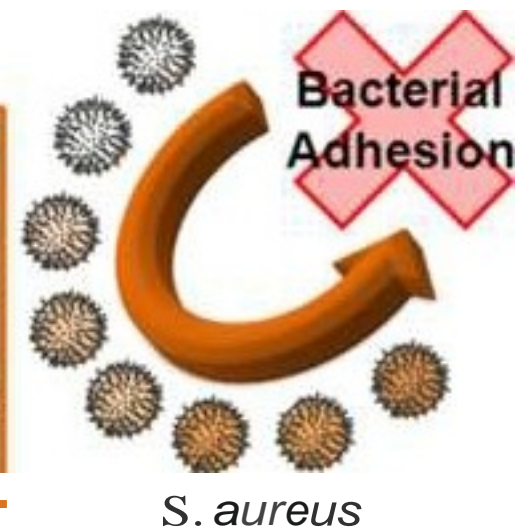